\documentclass[%
 reprint,
 amsmath,amssymb,
 aps,
]{revtex4-2}

\usepackage{graphicx}
\usepackage{dcolumn}
\usepackage{bm}
\usepackage{color}

\newcommand{\overbar}[1]{\mkern 1.5mu\overline{\mkern-1.5mu#1\mkern-1.5mu}\mkern 1.5mu}

\begin{document}

\preprint{APS/123-QED}

\title{Programmable skyrmion logic gates based on skyrmion tunneling  }

\author{Naveen Sisodia}
 \affiliation{Univ. Grenoble Alpes, CNRS, CEA, SPINTEC, F-38000 Grenoble, France}
 \author{Johan Pelloux-Prayer}
 \affiliation{Univ. Grenoble Alpes, CNRS, CEA, SPINTEC, F-38000 Grenoble, France}
 \author{Liliana D. Buda-Prejbeanu}
  \affiliation{Univ. Grenoble Alpes, CNRS, CEA, SPINTEC, F-38000 Grenoble, France}
 \author{Lorena Anghel}
 \affiliation{Univ. Grenoble Alpes, CNRS, CEA, SPINTEC, F-38000 Grenoble, France}
 \author{Gilles Gaudin}
 \affiliation{Univ. Grenoble Alpes, CNRS, CEA, SPINTEC, F-38000 Grenoble, France}
 \author{Olivier Boulle}
 \affiliation{Univ. Grenoble Alpes, CNRS, CEA, SPINTEC, F-38000 Grenoble, France}
 \email{olivier.boulle@cea.fr}

\date{\today}

\begin{abstract}

Magnetic skyrmions are promising candidates as elementary nanoscale bits in logic-in-memory devices, intrinsically merging high density memory and computing capabilities.  Here we  exploit  the dynamics of skyrmions interacting with  anisotropy energy barriers patterned by ion irradiation  to design programmable logic gates. Using micromagnetic simulations with experimental parameters, we show that a fine tuning of the barrier height and width allows the selective tunneling of skyrmions between parallel nanotracks triggered by skyrmion-skyrmion interaction. This can be leveraged to design skyrmion De-multiplexer (DMux) logic gate which works solely using skyrmions as logic inputs.  By cascading and connecting  demultiplexer   gates with a specific topology, we  develop a fully programmable logic gate capable of producing any possible logic output as a sum of all minterms generated by a given set of  inputs  without requiring any complex additional electric/magnetic interconversion.    The proposed design is fully conservative and cascadable and paves a new   pathway for   full  skyrmionic-based logic-in-memory devices.
\end{abstract}

\maketitle

\section{\label{sec:intro}Introduction}
The information technology industry is currently facing major challenges related to power dissipation and energy consumption~\cite{theis_end_2017,zhirnov_limits_2003}. The conventional Moore’s approach of CMOS technology is currently out of breadth: the continuously decreasing size of the MOS transistors in the logic and memory units leads to critical power dissipation and energy consumption issues. Another  major bottleneck relates to the Von Neumann architecture, in which the constant transfer of data between the memory and computing units  leads to a considerable cost in energy and limited bandwidth.  In addition, since the current technologies use volatile on-chip memory modules, the static power consumption used to maintain the stored data is significant even if state of the art low power strategies are used. Recently, logic-in-memory architectures, merging non-volatile memories and logic circuits, have attracted increased attention, as they are expected to realize ultra-low-power, higher bandwidth and shorten interconnection delays. Such an architecture opens the way for ``normally-off / instant–on" computing with no ``standby power" and wider memory bandwidth. This has led to the search for technologies that combine memory and computing capabilities in the same device.
 
Topological spin textures named magnetic skyrmions~\cite{nagaosa_topological_2013,FertReyrenCros_NRM2017} have emerged recently as a promising candidate to act as the building blocks of  logic-in-memory technologies that intrinsically merge high density non-volatile storage memories and logic operations. Magnetic skyrmions are local whirling of the magnetization. Their nanometric  scale, topological protection,  mutual repulsion, fast and low power manipulation can be exploited to code data and perform computation at the nanoscale. Different types of logic devices based on skyrmions have been proposed in recent years.~\cite{zhang2015magnetic,xing2016skyrmion,he2017current,luo2018reconfigurable,zhang2019skyrmion,chauwin2019skyrmion,song2021logic,zhang2020stochastic,gnoli2021skyrmion}. The primary physical phenomenon driving these devices is the mutual repulsion of skyrmions resulting from the dipolar and exchange interactions. Skyrmions are also very sensitive to local variations of the magnetic properties, which can be exploited to design local potential wells that guide the skyrmion trajectory in logic circuits~\cite{juge2021helium,songGuidingDynamicSkyrmions2020, purnamaGuidedCurrentinducedSkyrmion2015, laiImprovedRacetrackStructure2017, fookMitigationMagnusForce2015, toscanoSuppressionSkyrmionHall2020, bhattiEffectDzyaloshinskiiMoriya2019, loretoManipulationMagneticSkyrmions2019, angBilayerSkyrmionDynamics2019, iwasakiColossalSpinTransfer2014, laiMotionSkyrmionsWellSeparated2017, zhangSkyrmionskyrmionSkyrmionedgeRepulsions2015, yanRobustSkyrmionShift2020, angElectricalControlSkyrmion2020, sapozhnikovArtificialDenseLattice2016, toscanoBuildingTrapsSkyrmions2019, sapozhnikovMagneticSkyrmionsThicknessModulated2018}. 

As the anisotropy can be modulated using a gate electric field, programmable skyrmion switches and routers controlled by gate voltage have been proposed, which can be utilized to perform boolean logic  as well as  neuromorphic  and stochastic computing~\cite{pinna_skyrmion_2017,zhang2020stochastic}. However,  such an approach is plagued by the need for densely packed gates with bulky electrical circuitries and multiple voltage/skyrmion signal interconversion stages, resulting in a prohibitively large energy and footprint cost.

\begin{figure*}
\includegraphics[width=\linewidth]{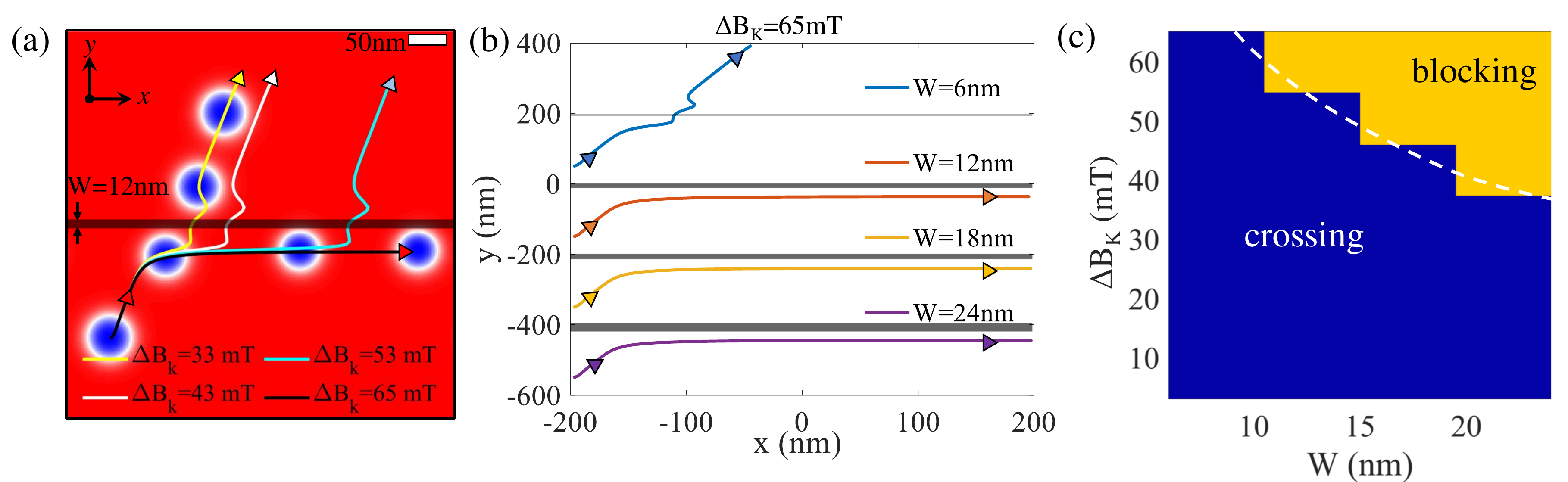}
\caption{\label{fig:barrier}(a) Trajectory of a skyrmion crossing an energy barrier created by a sharp change in anisotropy for different values of anisotropy difference $\Delta B_{\rm K}$ at fixed injected current density of $J=5\times10^{10}\rm A/m^2$.   The total size of the system is $\rm512nm\times512nm$. (b) Trajectory of a skyrmion for different widths of the barrier, $W$, at a fixed $\Delta B_{\rm K}=65\rm mT$ and current density $J=5\times10^{10}\rm A/m^2$. (c) $\Delta B_{\rm K}-W$ map for $J=5\times10^{10}\rm A/m^2$ indicating the set of parameter values for which the skyrmion can cross the barrier or is blocked by it. The dashed line is a guide to the eye indicating the boundary between both regions.}
\end{figure*}

 A more promising approach lies in a local modulation of the magnetic parameters through  engineering of the material. In particular, light ion irradiation is a powerful tool to  tailor the magnetic properties  of ultrathin films,  via  a gentle intermixing of the interfaces~\cite{chappert_planar_1998}. In ultrathin Pt/Co/MgO films, we have shown that it leads to a decrease of the magnetic anisotropy and Dzyaloshinkii-Moriya interaction~\cite{juge2021helium}.  This can  be exploited to create and guide skyrmions in   racetracks defined by ion irradiation, which acts as local potential wells for the skyrmions~\cite{juge2021helium}.

In this work,  we   investigate the interactions of skyrmions  with anisotropy energy barriers and show that  a fine tuning of the barrier height and width  allows the selective tunneling of  skyrmions between parallel nanotracks triggered by skyrmion-skyrmion interaction. This can be leveraged  to design a skyrmion De-multiplexer (DMux) logic gate which works solely using skyrmions as logic inputs. By cascading three demultiplexer logic gates, a fully programmable logic gate is proposed which allows the realization of any desired logic output from a set of given inputs. The developed logic design is fully conservative, i.e., no skyrmion is destroyed during the process and all skyrmions can be recovered at the end of the logic operation, which makes it suitable for conservative reversible logic. This also removes the need for continuous generation of skyrmions which is an energy expensive process. The developed logic design can significantly reduce the number of elementary logic operations required to perform complex arithmetic computations. Moreover, the proposed implementation also provides a modular design template for creating large scale logic networks by re-purposing DMux gates.

\section{\label{sec:barrdesign} Skyrmion dynamics in presence of an energy barrier}
In this section, we establish the basic understanding of the dynamics of a single skyrmion interacting with an energy barrier created by a sudden change in anisotropy and DMI induced by He-ion irradiation. 
We performed micromagnetic simulations with magnetic and transport parameters consistent with our prior experimental work (see Appendix \ref{sec:methodsmicro} for details regarding the micromagnetic model and parameters)~\cite{juge2021helium}. 
 
\begin{figure*}
\includegraphics[width=\linewidth]{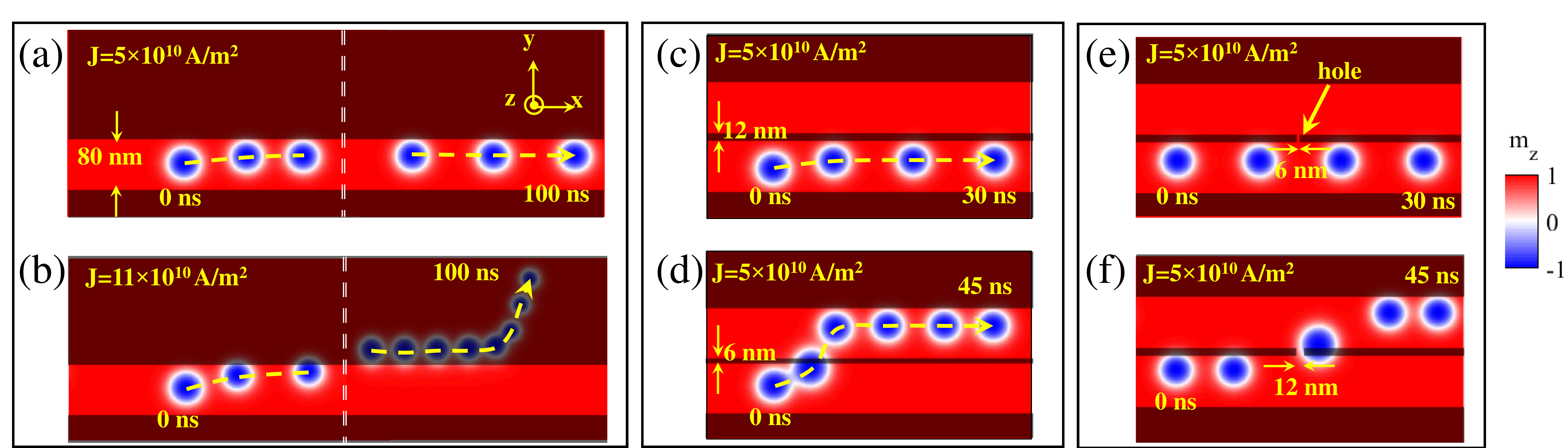}
\caption{\label{fig:basic} Skyrmion motion in   $80\rm nm$ wide nanotrack for current density (a) $J=5\times10^{10}\rm A/m^2$ and (b) $J=11\times10^{10}\rm A/m^2$, respectively. The dashed white lines on the track  represent a discontinuity along the x-scale which we use to show the motion for a longer time-scale ($t=0-100$ns). The skyrmion has a  diameter $\sim 57\rm nm$. Skyrmion motion in two parallel nanotracks separated by a non-irradiated barrier region of width $W=12\rm nm$ (c), and $W=6\rm nm$ (d), respectively, for a fixed current density of $J=5\times10^{10}\rm A/m^2$. Skyrmion motion in two parallel nanotracks separated by a barrier region of width, $W=12\rm nm$ having an inter-connecting channel of width $\rm6nm$ (e) and $\rm12nm$ (f), respectively, at the center of the barrier. Movies corresponding to all cases (a)-(f) are shown in supplementary video SV2}
\end{figure*}

We considered an ultrathin film with a single skyrmion moving under the influence of  spin orbit torque (SOT). An energy barrier of width $W$ and height $\Delta B_{\rm K}$ is present in the system, located at $y=0$. The barrier height $\Delta B_{\rm K}$  is defined as the difference between the values of the effective anisotropy field of the barrier (non-irradiated) and the outer (irradiated) region, $\Delta B_{\rm K}=B_{\rm K,non-irr}-B_{\rm K,irr}$. The skyrmion moves along an oblique direction at an angle of $\sim68^\circ$ (skyrmion Hall angle) with the x-axis due to the current flowing along $+x$-direction (current density, $J=5\times10^{10}\rm A/m^2$). Figure~\ref{fig:barrier}~(a) shows the trajectories of the skyrmion for different values of the barrier height $\Delta B_{\rm K}$ but with a fixed barrier width of $W=12\rm nm$. The anisotropy of the outer region is fixed at $B_{\rm K,irr}=7\rm mT$.   We find that  for low $\Delta B_{\rm K}$, the skyrmion can pass through the energy barrier while for large enough $\Delta B_{\rm K}$, the skyrmion does not pass and eventually moves along the barrier (see $\Delta B_{\rm K}=65\rm mT$). Interestingly, when the skyrmion   passes through the barrier, we observe that the trajectories are  shifted along the $x$-direction, the magnitude of this shift  being dependent on the value of $\Delta B_{\rm K}$. Similar behavior was also observed for skyrmions crossing energy steps~\cite{menezes_deflection_2019,castell2019accelerating}. The possibility of crossing the barrier also depends on the barrier width $W$.  We show in Fig.~\ref{fig:barrier}~(b), the trajectories of skyrmions for different  $W$  with a fixed barrier height of $\Delta B_{\rm K}=65\rm mT$. For low width ($W=6\rm nm$), the skyrmion easily crosses the barrier while for all other cases ($W=12,~18$ and $24\rm nm$), the skyrmion is blocked and moves along the barrier (see supplementary video SV1).  Physically, this can be understood by considering the fact that, in the case of a narrow barrier, only a small portion of the skyrmion falls inside the energetically unfavourable barrier region during the crossing from the lower to the upper region. Thus, the energy increase due to the skyrmion moving inside the non-irradiated region is not very high and can be overcome by the force due to the skyrmion Hall effect. We perform similar simulations for a range of $\Delta B_{\rm K}$ and $W$ and present a corresponding map in Fig.~\ref{fig:barrier}~(c) depicting the set of $\Delta B_{\rm K}$-$W$ values for which the skyrmion can either cross the barrier (low $\Delta B_{\rm K}$ and $W$) or be blocked by it (high $\Delta B_{\rm K}$ and $W$). The boundary between both regimes depends on  the injected  current: for a higher value of the current density, we expect the boundary to shift toward higher $\Delta B_{\rm K}$ and $W$, as the skyrmion has more energy to cross the energy barrier.

\subsection{\label{sec:trackdesign} Skyrmion motion inside irradiated nanotracks}
The possibility for a skyrmion to cross or not an energy barrier depending on the external parameters can be exploited to guide and control the skyrmion trajectory in tracks defined by ion irradiation. Figures~\ref{fig:basic}~(a) and (b) show the trajectory of a skyrmion in a nanotrack with lower anisotropy driven by a current density of $J=5\times10^{10}\rm A/m^2$ and $J=11\times10^{10}\rm A/m^2$, respectively. For a fixed $\rm \Delta B_{\rm K}$, there exists a threshold current density beyond which the skyrmion cannot be confined inside the track. For our simulation parameters, a  threshold $J_{\rm th}=10\times10^{10}\rm A/m^2$ is found. In the following, we use a current density of $5\times10^{10}\rm A/m^2$ which is half of this threshold, such  that the   skyrmion stays confined in the track.

Next, we consider the motion of skyrmions in parallel nanotracks. To minimize the area of the device which is critical for energy consumption, it is necessary to densely pack many tracks. In Fig.~\ref{fig:basic}~(c) and (d), we show the motion of a skyrmion in two parallel tracks separated by a barrier of width $W=12\rm nm$ and $W=6\rm nm$, respectively, for a current density of $J=5\times10^{10}\rm A/m^2$. The results are similar to Fig.~\ref{fig:barrier}(b) indicating that there is a minimum barrier width ($W_0$) which has to be maintained to ensure that the skyrmion stays confined in the track. For our parameters, this minimum barrier width is $W_0=10\rm nm$. For the simulations in the subsequent sections, we maintain a gap width $W=12\rm nm$ which is larger than $W_0$. We note here that the state-of-art FIB systems can easily produce sub-10nm features using He-ions~\cite{lewis2019plasma}.

The crossing between parallel tracks can also be controlled by adding an additional channel of finite width between the two tracks (see Fig.~\ref{fig:basic}~(e) and (f)). This channel has the same magnetic parameters as that of the irradiated nanotracks and it works as a  tunnel between the upper and the lower track. For a channel width of $6\rm nm$ [Fig.~\ref{fig:basic}~(e)], we find that the trajectory of the skyrmion is unaffected. However, for a channel width of $12\rm nm$ [Fig.~\ref{fig:basic}~(f)], the skyrmion can enter the upper track after traversing through the channel. Thus, even though the channel width for Fig.~\ref{fig:basic}~(f) is nearly 5 times lower than the skyrmion diameter, it is still enough for the skyrmion to pass through. A similar situation in a physically milled nanotrack may not be  achieved as the skyrmion would have to compress significantly to move through the channel during which it may be destroyed. This absence of any physical discontinuity in the system is an important feature of our simulations preventing  undesired destruction of skyrmion at  physical edges.

\begin{figure*}
\includegraphics[width=\linewidth]{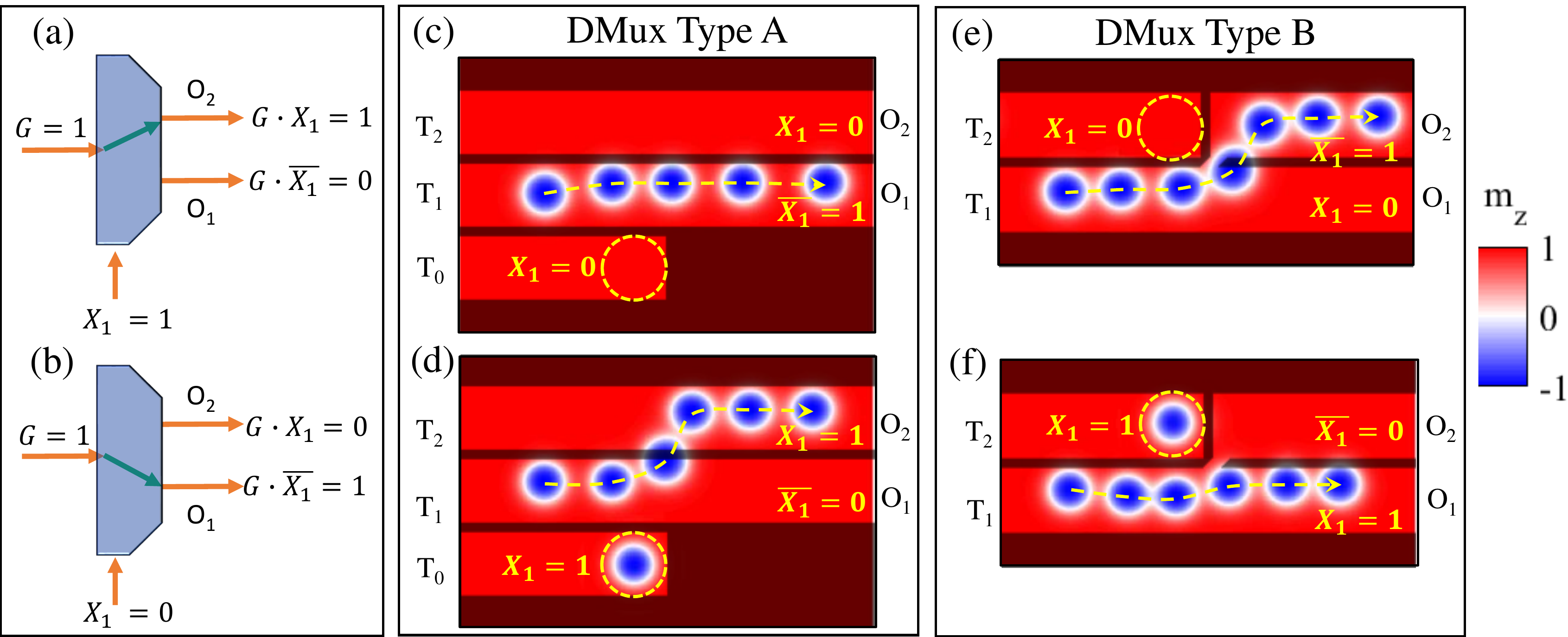}
\caption{\label{fig:dmuxtype}(a) and (b) show the schematic and operation of a 1-to-2 DMux gate which produces a 0(1) output at $O_1$ and 1(0) at $O_2$ if the selector input is $X_1=1(0)$. The input ``G" is assumed to be fixed at ``1". (c) and (d) show the micromagnetic simulation of a skyrmion based DMux gate where the incoming signal ``G" is a skyrmion in track $T_1$. The selector input is a skyrmion (or no-skyrmion) in track $T_0$. The current density is $J=5\times 10^{10}\rm A/m^2$. (e) and (f) shows the micromagnetic simulations of a different design implementation of skyrmion based DMux with an additional channel in the barrier. For this case, the track for selector input is above the ``G" input. Movies corresponding to (c)-(f) are shown in supplementary video SV3}
\end{figure*}

\subsection{\label{sec:dmuxmain}Skyrmion based De-multiplexer}
The concepts of skyrmion movement within and passing between parallel nano-tracks can be exploited to design a skyrmion-based De-multiplexer (DMux) logic gate. A DMux gate is a logic gate which connects an incoming signal to one of the multiple output streams based on the value of \textit{selector} input. We will here discuss a 1-to-2 DMux which takes a single \textit{selector} ($X_1$) input (``0" or ``1") and connects the incoming signal ($G$) to one of the two outputs ($O_1$ or $O_2$) (see Fig.~\ref{fig:dmuxtype}~(a) and (b)). In our device, the input ``1” is represented by a single skyrmion, while the absence of skyrmion will correspond to input ``0”. Two different DMux gate designs are presented, both of which are based on skyrmions moving inside parallel nanotracks but differ in their working principle. These  two different DMux gates can be connected to build programmable logic gates as will be shown later in Sec.~\ref{sec:dmuxfull}.  

\subsubsection*{\label{sec:dmuxtypeA}De-multiplexer : Type A}

 Figures~\ref{fig:dmuxtype}(c) and (d) show the working principle of our first DMux gate design, named ``DMux \textit{Type A}". The width of each track is $\rm80nm$ with a barrier  width $W=12\rm nm$ separating each track. The incoming skyrmion enters the system from the left-hand side of the $T_1$ track (``G" input) while the selector input ($X_1$) is in the $T_0$ track. A uniform DC current along the $\hat{x}$-direction ($J=5\times 10^{10}\rm A/m^2$)  moves the skyrmion with a positive velocity $v_x$ in the $\hat{x}$-direction as discussed previously. For the case where the selector input “$X_1$” is “0”[Fig.~\ref{fig:dmuxtype}(c)], i.e., there is no skyrmion in the $T_0$ track, the skyrmion input ``G" moves along the $T_1$ track leading  to the output $O_1=1$. The output $O_2$ remains ``0" as no skyrmion is present in the corresponding track. The Fig.~\ref{fig:dmuxtype}(d) shows the operation of DMux if the selector input is “1”, i.e., a skyrmion is present in the $T_0$ track. This skyrmion in the track $T_0$ cannot move along the $+x$-direction during the current injection as it is blocked by the energy barrier. When the skyrmion in the track $T_1$ gets closer to the skyrmion in the track $T_0$, which remains almost fixed due to the barrier, it experiences a repulsive force due to the skyrmion-skyrmion interaction. This repulsive force enhances the transverse force already acting on the skyrmion in track $T_1$ due to the skyrmion Hall effect. The resulting force is enough to push the moving skyrmion through the barrier between track $T_1$ and $T_2$, and this skyrmion ends its trajectory at output  $O_2$. From the results in Fig.~\ref{fig:dmuxtype}(c) and (d), we conclude that the overall functionality of DMux is achieved:   the incoming skyrmion from the left of the $T_1$ track (``G") is guided to   either the output $O_1$ or the output $O_2$ depending upon the presence of skyrmion in the $T_0$ track (\textit{selector} input). The outputs can be expressed as : $O_1=\overbar{X_1}$ and $O_2=X_1$ (for $G=1$).

\subsubsection*{\label{sec:dmuxtypeB}De-multiplexer : Type B}
In the design of DMux \textit{Type A}, the track for the \textit{selector} input $X_1$ is below the track for the ``G" input. In our programmable logic gate design, we will also require a DMux gate in which the moving skyrmion (“G” input) is below the \textit{selector} input ($X_1$). The principle of this DMux \textit{Type B} design functionality is shown in Fig.~\ref{fig:dmuxtype}(e) and (f). Similar to our previous design, we use irradiated tracks to confine the skyrmion, however, we remove the track $T_0$ where the input $X_1$ was kept in DMux \textit{Type A}. Instead, we divide the track $T_2$ in two parts by adding a vertical barrier of width $W=12\rm nm$ in the middle of the $T_2$ track. The input $X_1$ is now sent from the left of the track $T_2$.
Additionally, we connect the track $T_1$ with the right part of $T_2$ by introducing a small channel in the barrier, which allows a skyrmion moving in track $T_1$ to pass to the track $T_2$. In Fig.~\ref{fig:dmuxtype}(e), the incoming skyrmion enters from the left of $T_1$ track and moves along $+x$-direction. If there is no skyrmion in the upper track ($X_1=0$), the skyrmion will move towards the upper track $T_2$ tunneling through the interconnecting channel and will arrive at the output $O_2$. However, if the track $T_2$ has a skyrmion ($X_1=1$) [Fig.~\ref{fig:dmuxtype}(f)], the skyrmion in the track $T_1$ moves straight to the right end towards the output $O_1$ due to the mutual repulsive force. Overall, as in the previous case, the DMux logic here is achieved as the skyrmion is guided to the right side at either output $O_1$ or output $O_2$ depending on the selector input ($X_1$). The output of DMux \textit{Type B} can be expressed as : $O_1=X_1$ and $O_2=\overbar{X_1}$ (for $G=1$).

\subsection{\label{sec:dmuxfull}Programmable logic gate design using De-multiplexers}

\begin{figure}
\includegraphics[width=\linewidth]{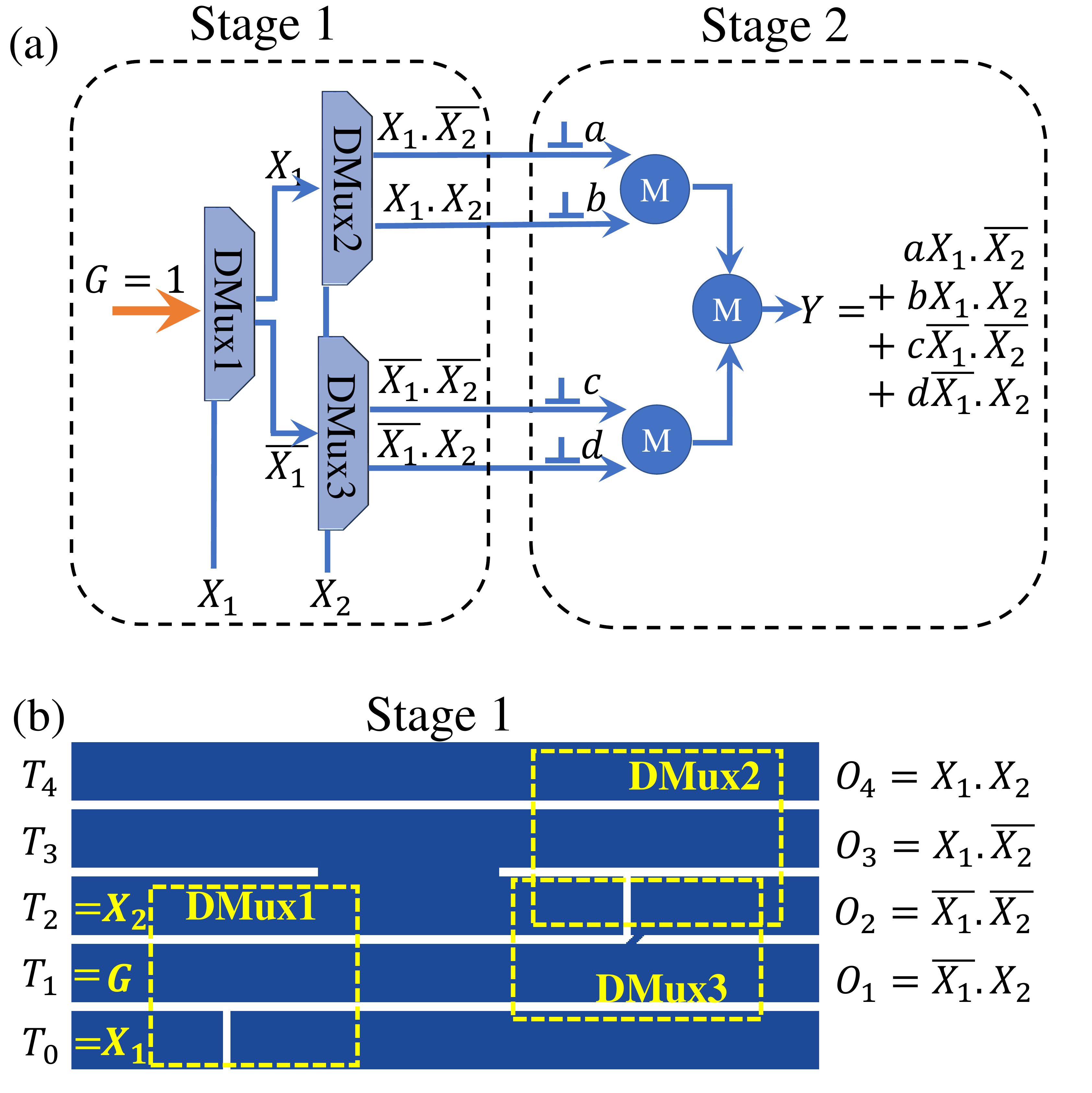}
\caption{\label{fig:reconfigdesign}(a) Schematic of a programmable logic gate design utilizing three DMux gates. (b) shows the design of programmable logic (\textit{Stage 1}) using two DMux \textit{Type A} logic gates (DMux1 and DMux2) and one DMux \textit{Type B} logic gate (DMux3).}
\end{figure}

De-multiplexers are extremely useful logic gates and can be connected in several different ways to perform a variety of logic operations. Such an implementation was discussed in Ref.~\cite{kluge1972computation} to achieve a programmable logic gate  based on  three DMux gates which can produce many logical combinations of the two input signals (see Fig.~\ref{fig:reconfigdesign}(a)). For clarity, we split the entire operation into two stages. The \textit{Stage 1} involves two input bits $X_1$ and $X_2$ which are passed as selection inputs for different DMux gates. The operation flow is as follows: $X_1$ is passed to the first DMux gate (DMux1) which outputs either $X_1$ or $\overbar{X_1}$ for an incoming signal $G=1$ (fixed and constant).
The outputs of DMux1 act as the ``G" input for the next gates DMux2 and DMux3. The selection input of both these gates is set using $X_2$. In the end, we obtain all four possible minterms with $X_1$ and $X_2$, namely $X_1.\overbar{X_2}$, $X_1.X_2$, $\overbar{X_1}.\overbar{X_2}$ and $\overbar{X_1}.X_2$.

In the \textit{Stage 2},  switch gates [$a,b,c,d=(0,1)$]  are used on each of the four output bits coming from DMux2 and DMux3 which are then added to obtain the final output:
$Y=a(X_1.\overbar{X_2})+b(X_1.X_2)+c(\overbar{X_1}.\overbar{X_2})+d(\overbar{X_1}.X_2)$.
By setting the [$a,b,c,d$] values, any logical operation on  $X_1$ and $X_2$ can be achieved. 

Designing this programmable gate with skyrmion based DMux gates involves that (i) the input $X_2$ (a skyrmion) must be conserved (stored) or duplicated and  pass through both DMux2 and DMux3 gates and (ii) the output of the first DMux gate has to reach the input of the next DMux gate (cascading) without altering its value. 
We first focus on \textit{Stage 1} of this logic circuit and propose a design in Fig.~\ref{fig:reconfigdesign}(b), which is based on the previously described DMux gates \textit{Type A} and \textit{Type B}. The tracks in DMux2 and DMux3 which carry the selection input (track $T_2$) are merged to avoid the need for duplication of input $X_2$. This is possible only if we use the \textit{Type A} design for the DMux2 and \textit{Type B} design for the DMux3 as shown in Fig.~\ref{fig:reconfigdesign}(b). This merging also reduces the total number of required tracks, reducing the total area of the device, which further minimizes the energy consumption for the operation.

\subsubsection*{\label{stage1}Performing Logic operations (Stage 1)}
\begin{figure*}
\includegraphics[width=\linewidth]{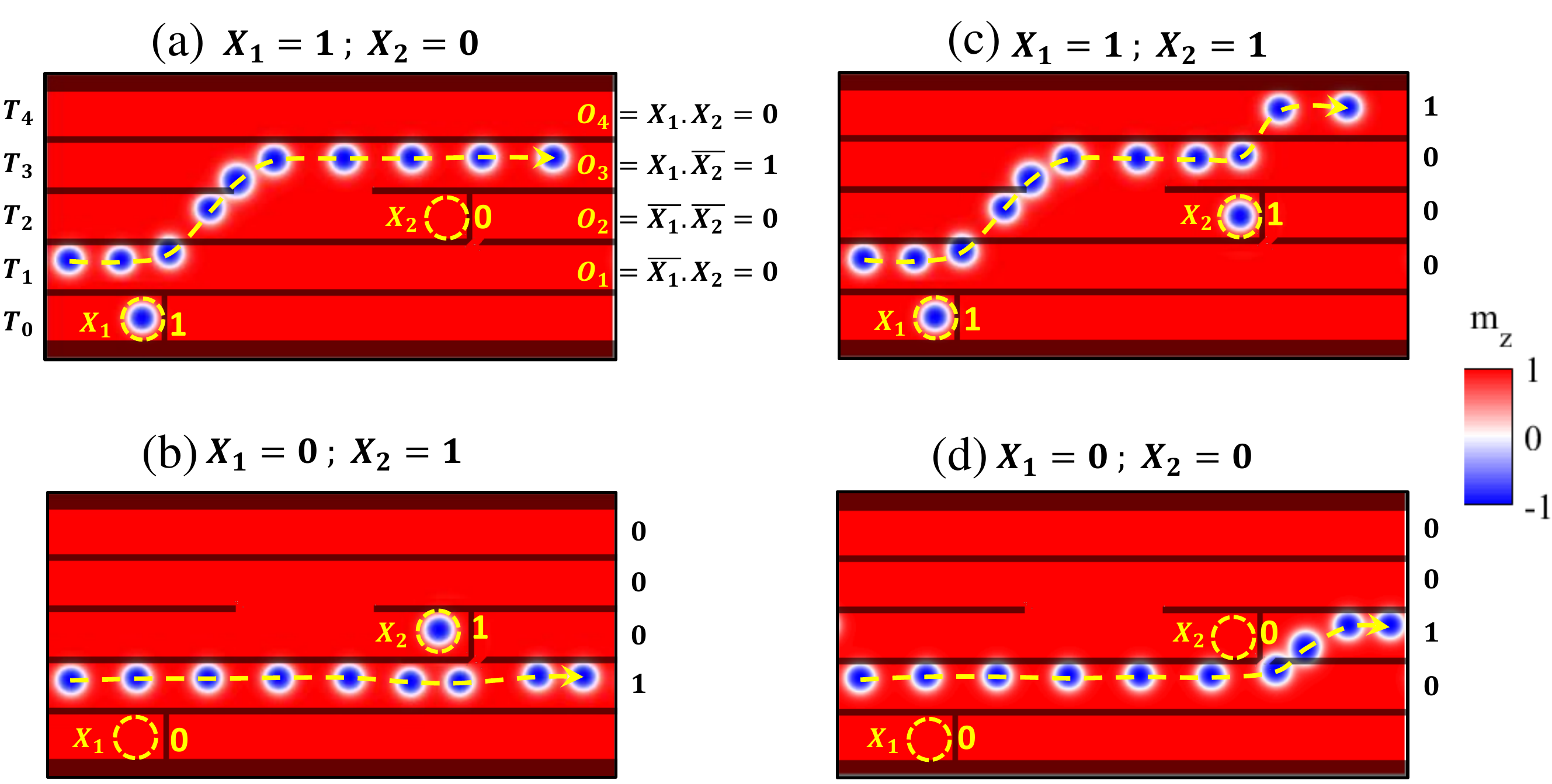}
\caption{\label{fig:threedmux}Micromagnetic simulation of the  programmable logic \textit{Stage 1} showing the trajectory of the skyrmion for various input cases. Depending on the inputs $X_1$ and $X_2$, the skyrmion ``G" entering through track $T_1$ reaches one of the outputs $O_1-O_4$. Note that for each set of inputs $X_1$ and $X_2$, only one of the output tracks ($O_1-O_4$) has a skyrmion (``1") and the rest are ``0".}
\end{figure*}

Figure~\ref{fig:threedmux} shows the operation of our programmable logic gate design proposed in Fig.~\ref{fig:reconfigdesign}~(b) using micromagnetic simulations. Here, we assume that the inputs $X_1$ and $X_2$ are already correctly placed in the tracks $T_0$ and $T_2$ at their respective positions. The exact protocol to place these inputs is discussed in Appendix~\ref{sec:inputplacement}.
For the case of $X_1=1$ and $X_2=0$ [Fig.~\ref{fig:threedmux}(a)], the skyrmion input ``G" enters through track $T_1$ and is pushed to track $T_2$ due to the repulsion from the skyrmion in track $T_0$ (input $X_1$). The skyrmion then passes through the large opening in the barrier into track $T_3$. In track $T_3$, the skyrmion keeps moving straight (along $+x$-direction) and is obtained at the output end $O_3$. For the case of $X_1=0$ and $X_2=1$ [Fig.~\ref{fig:threedmux}(b)], the skyrmion enters through track $T_1$ and initially moves straight without facing any repulsion (as input $X_1$ is "0"). The skyrmion then experiences repulsion from the skyrmion in track $T_2$ (input $X_2$). Due to this repulsion, the moving skyrmion cannot tunnel through the channel between tracks $T_1$ and $T_2$ 
and is forced to move in the same track $T_1$, to finally exit through the output end $O_1$. For inputs $X_1=1$ and $X_2=1$ [Fig.~\ref{fig:threedmux}(c)], the input skyrmion ``G" follows the same pattern as in Fig.~\ref{fig:threedmux}(a) till it reaches the track $T_3$. While moving in the track $T_3$, it is repelled from the skyrmion input $X_2$ lying in the track $T_2$ which pushes it to the uppermost track $T_4$, \emph{i.e.} output $O_4$ (see also supplementary video SV4). Lastly, for the case of $X_1=0$ and $X_2=0$, the input skyrmion ``G" in the track $T_1$ moves straight along the track till it reaches the interconnecting channel between tracks $T_1$ and $T_2$. After tunneling through the channel to the $T_2$ track, the skyrmion then reaches the $O_2$ output. 
Using the previous expressions for \textit{Type A} and \textit{Type B} DMux gates, we can write the mathematical notation for each of the output gates $O_1-O_4$ as follows: $O_1=(\overbar{X_1}.X_2), ~O_2=(\overbar{X_1}.\overbar{X_2}), ~O_3=(X_1.\overbar{X_2})$ and $O_4=(X_1.X_2)$. These outputs can now be combined in the \textit{Stage 2} operation of the logic circuit for obtaining the desired operation as discussed in the next section.

\begin{figure}
\includegraphics[width=\linewidth]{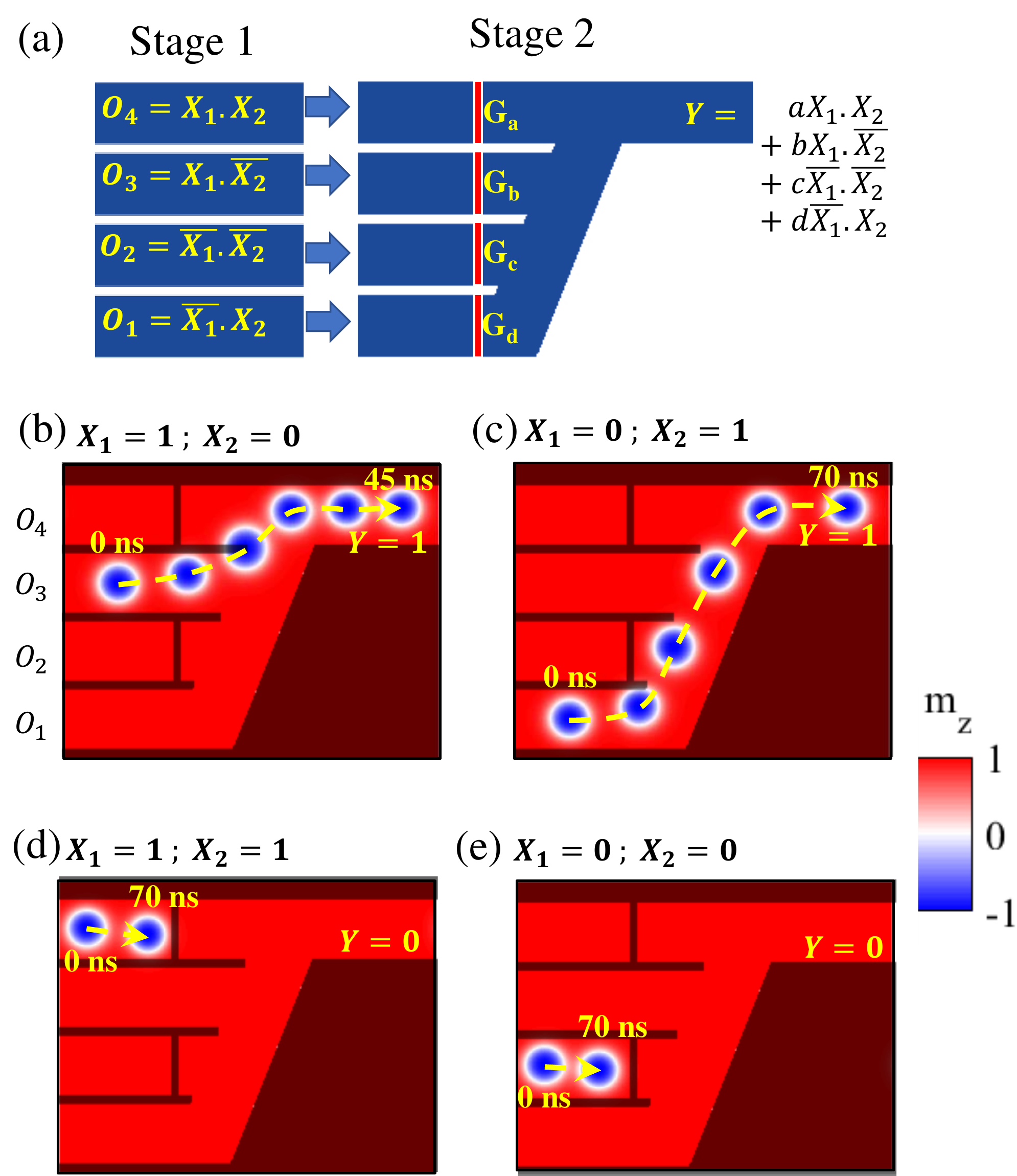}
\caption{\label{fig:stage2micro}(a) Schematic of the \textit{Stage 2} design of the programmable logic gate shown in Fig.~\ref{fig:reconfigdesign}~(a) using irradiated nanotracks. The inputs of the \textit{Stage 2} are the outputs from the \textit{Stage 1} operation shown in Fig.~\ref{fig:threedmux}. Gates $G_{a}-G_{d}$ control the final output $Y$. 
(b)-(e) show the micromagnetic simulations of the \textit{Stage 2} operation in programmable logic for achieving the output of an $\rm \textbf{XOR}$ gate for different input cases. Gates $G_a$ and  $G_c$ are used (non-irradiated regions) to perform the required logic operation : $Y=X_1.\overbar{X_2}+\overbar{X_1}.X_2$.}
\end{figure}

\subsubsection*{\label{stage2}Combining Outputs (Stage 2)}

The \textit{Stage 2} of Fig.~\ref{fig:reconfigdesign}~(a) involves passing the four outputs from \textit{Stage 1} through an ON/OFF gate and combining them afterward. These ON/OFF gates can simply be constructed as tracks with/without an energy barrier in the track which prevents/allows the skyrmion to pass. Note that the programmability in this design is limited to modifying the tracks to obtain any desired logic functionality which is known at the time of device fabrication. However, a dynamically reconfigurable design can also be achieved by replacing these barriers with voltage-controlled gates which can modify the magnetic anisotropy locally and can thus reconfigure the logic functionality even after fabrication. The output obtained after each of these gates is finally merged to obtain the final output $Y$. This merging is relatively easy to implement in our design by simply adding another track at an angle to the parallel tracks combining all four output tracks ($O_1-O_4$). For faster skyrmion motion, the angle between the new track and the four parallel tracks is kept at $68^\circ$ which is the same as the Skyrmion Hall Angle. The full design is shown in Fig.~\ref{fig:stage2micro}(a) where the regions $G_a-G_d$ represent the gates and the final output $Y$ is obtained from the uppermost track. It may be noted here that for any combination of inputs $X_1$ and $X_2$ only one of the outputs $O_1-O_4$ contains a skyrmion (``1") and the rest are empty (``0"). Therefore, there is no possibility of multiple skyrmions arriving from two different tracks and interacting with each other during the merging operation through the non-parallel track.

To demonstrate the working principle of this design using micromagnetic simulations, we use the expression for $\rm \textbf{XOR}$ gate as an example of our desired output. Other logic functionalities can also be obtained by changing the values of parameters [a,b,c,d] as given in Table~\ref{tab:stage2param}. The $\rm \textbf{XOR}$ gate is mathematically represented as : $Y=X_1.\overbar{X_2}+\overbar{X_1}.X_2$. 
To implement the \textbf{XOR} logic operation, only the gates $G_a$ and $G_c$ work as energy barriers (non-irradiated regions) in tracks corresponding to $O_4$ and $O_2$. The micromagnetic simulations for all four combinations of input $X_1$ and $X_2$ are shown in Fig.~\ref{fig:stage2micro}(b)-(e). For the case ($X_1=1$, $X_2=0$), [Fig.~\ref{fig:stage2micro}~(b)] and ($X_1=0$, $X_2=1$), [Fig.~\ref{fig:stage2micro}~(c) and supplementary video SV5]  the output skyrmion of \textit{Stage 1}  proceeds to the uppermost output track giving the final output $Y=1$. For the case of ($X_1=1, X_2=1$) [Fig.~\ref{fig:stage2micro}~(d)] and ($X_1=0, X_2=0$) [Fig.~\ref{fig:stage2micro}~(e)], the output from \textit{Stage 1} is received in $O_4$ and $O_2$, respectively, both of which are blocked by the energy barriers. Thus, the final output for both these cases is $Y=0$.

\begin{table}[b]
\caption{\label{tab:stage2param}%
Values of parameters (a,b,c,d) for obtaining different logic operations. The output is represented by $Y=a(X_1.X_2)+b(X_1.\overbar{X_2})+c(\overbar{X_1}.\overbar{X_2})+d(\overbar{X_1}.X_2)$
}
\begin{ruledtabular}
\begin{tabular}{lcccc}
\textrm{Logic Gate}&
\textrm{a}&
\textrm{b}&
\textrm{c}&
\textrm{d}\\
\colrule
AND & 1 & 0 & 0 & 0\\
OR & 1 & 1 & 0 & 1\\
XOR & 0 & 1 & 0 & 1\\
NAND & 0 & 1 & 1 & 1\\
NOR & 0 & 0 & 1 & 0\\
XNOR & 1 & 0 & 1 & 0\\
\end{tabular}
\end{ruledtabular}
\end{table}

\subsubsection*{\label{energy}Operational Characteristics}

For the material parameters and dimensions used in the micromagnetic simulations, the operation time for the designed gate varies from $76-120\rm~ns$ for \textit{Stage 1} and $45-70\rm~ns$ for \textit{Stage 2} with a current of 0.1~mA which corresponds to a maximum energy dissipation of $238\rm~fJ$ (\textit{Stage 1}+\textit{Stage 2}) by Joule heating. Note that the  magnetic and transport parameters of  the material stacks were extracted from proof of concept experiments, which were not designed for logic gates implementations. Faster operations with lower energy can be easily achieved by designing the gate with magnetic materials with lower damping, larger spin orbit torques,  higher anisotropy in the non-irradiated track and lower device dimensions. For instance, by considering materials with 10 nm skyrmions, a magnetic damping of 0.05 and assuming the lateral dimensions decreased by a factor 5 due to smaller skyrmion size,  a total operation time of $6.3$ ns with a current of $20\rm~\mu A$ and an energy dissipation of $0.32$~fJ is anticipated for $J=5\times10^{10}\rm~A/m^2$. Note that these values do not include the delay and energy dissipated in the addressing transistors and contact lines.   These figures should be compared with the  energy dissipation and delay in interconnects for the data to reach the CMOS logic in a conventional Von Neumann architecture, which is typically of the order of the pJ for 1mm interconnect, combined with the  write time in SRAM cache memory, around the ns. Thus, we expect a drastic decrease in the energy consumption as compared to a standard Von Neumann architecture.

\section{\label{sec:Conclusions}Conclusions}
To conclude, we have proposed a programmable logic gate design based on skyrmion-skyrmion interactions. The gate exploits the skyrmion guiding and selective crossing of energy barrier designed by local patterning of the magnetic parameters, which can be realized using light ion-irradiation.  The gate is conservative and cascadable and since it relies purely on skyrmion interactions, does not require complex electric/magnetic interconversion gates. Since skyrmions can be used at nanoscale to represent elementary bits, the proposed gate could form the basis of logic-in-memory devices that intrinsically merge high density memory and computing capabilities.

\begin{acknowledgments}
The authors acknowledge financial support from the French national research agency (ANR) (Grant Nos. ANR-15-CE24-0015-01 and ANR-17-CE24-0045) and the American defense advanced research project agency (DARPA) TEE program (Grant No. MIPR HR0011831554). This work has been partially supported by MIAI@Grenoble Alpes, (ANR-19-P3IA-0003).
\end{acknowledgments}

\appendix
\section{\label{sec:methodsmicro}Micromagnetic simulations}
The simulations are performed  in a micromagnetic framework by numerically solving the Landau-Lifshitz-Gilbert (LLG) equation using the open-source Mumax3 package~\cite{VansteenkisteWaeyenberge_AIPadv2014}. 
We consider a stack composed of ferromagnetic/heavy metal (FM/HM)   ultrathin films. Due to the spin-orbit interaction at the ferromagnetic/heavy metal (FM/HM) interface, a spin current flows in the perpendicular ($+\hat{z}$) direction through the FM if a charge current is injected in the in-plane $+\hat{x}$-direction in the HM. The direction of polarization of the electrons $\textbf{e}_{\rm p}$ is along the $-\hat{y}$-direction, \emph{i.e}  orthogonal to the direction of flow of charge current as well as the flow of spin current. We include the additional torque on the FM due to this spin current using the Slonczewski model.

\begin{gather*}\label{eq:llgs}
\frac{d\textbf{m}}{dt}=-\gamma (\textbf{m}\times \textbf{B}_{\rm eff}) + \alpha(\textbf{m}\times \frac{d\textbf{m}}{dt})\\
-\mathcal{T} \Big(\textbf{m}\times(\textbf{m}\times\textbf{e}_{\rm p})+\epsilon^{'}(\textbf{m}\times\textbf{e}_{\rm p}) \Big)\\
\mathcal{T}=\frac{\hbar\theta_{\rm H} J_{\rm c} }{2|e|M_{\rm S}d}
\end{gather*}

where, $\bf{m}$ is the normalized local magnetic moment, $\gamma$ is the gyromagnetic ratio of the electron, $\alpha$ is the Gilbert damping constant and $\mathcal{T}$ represents the strength of spin orbit torque. $J$ is the charge current through the HM flowing in $+\hat{x}$-direction, $|e|$ is the electronic charge, $M_{\rm S}$ is the saturation magnetization and $d$ is the thickness of the ferromagnetic film. We use $\theta_H=0.1$ as the spin Hall angle and $\epsilon^{'}=0.015$ as the ratio of field-like to damping-like SOT torque.
$\bf{B}_{\rm eff}$ is the effective field which includes contributions from the external field ($\bf{B}_{\rm ext}$), the magnetostatic interactions ($\bf{B}_{\rm m}$), the exchange interactions ($\bf{B}_{\rm exch}$), the perpendicular magnetic anisotropy ($\bf{B}_{\rm an}$) and Dzyaloshinskii Moriya interaction ($\bf{B}_{\rm DMI}$).

To achieve realistic predictions, we use the material parameters corresponding to our previous experimental work on skyrmion confinement in He irradiated nanotracks~\cite{juge2021helium} given in Table~\ref{tab:parameter} with an external field of $\textbf{B}_{\rm ext}=33\rm mT~\hat{z}$. Note that we obtain the value of effective anisotropy field ($B_{K}$) from experiments which can be translated to the uniaxial anisotropy constant using the expression: $K=(B_{K}M_{S}/2)+(\mu_{0}M_{S}^2/2)$.
The total area of the simulated system in \textit{Stage 1} is $\rm1024nm\times 512nm$ with $\rm0.9nm$ FM thickness. The area of \textit{Stage 2} design is $\rm512nm\times 512nm$. The resistivity of the HM (Pt in our case) is $30 \Omega\mu m$ from experimental measurements.
For solving the LLG equation, we discretize the entire FM into cuboidal cells of $\rm 2nm\times2nm\times0.9nm$ and use the $4^{th}$ order Runge-Kutta method with an adaptive time-step.

\begin{table}[b]
\caption{\label{tab:parameter}%
Simulation Parameters
}
\begin{ruledtabular}
\begin{tabular}{lcc}
\textrm{Parameter}&
\textrm{Irradiated}&
\textrm{Non-irradiated}\\
\colrule
$M_s~\rm (MA/m)$ & 1.32 & 1.32\\
$\mu_0 B_{\rm K}~\rm (mT)$ & 7 & 72\\
{$\vert{}D\vert~\rm (mJ/m^2)$} & 1.082 & 1.122\\
$A\rm~(pJ/m)$ & 16 & 16\\
${\alpha}$ & 0.3 & 0.3\\
\end{tabular}
\end{ruledtabular}
\end{table}

\section{\label{sec:inputplacement}Setting and recovering inputs}

\begin{figure}
\includegraphics[width=\linewidth]{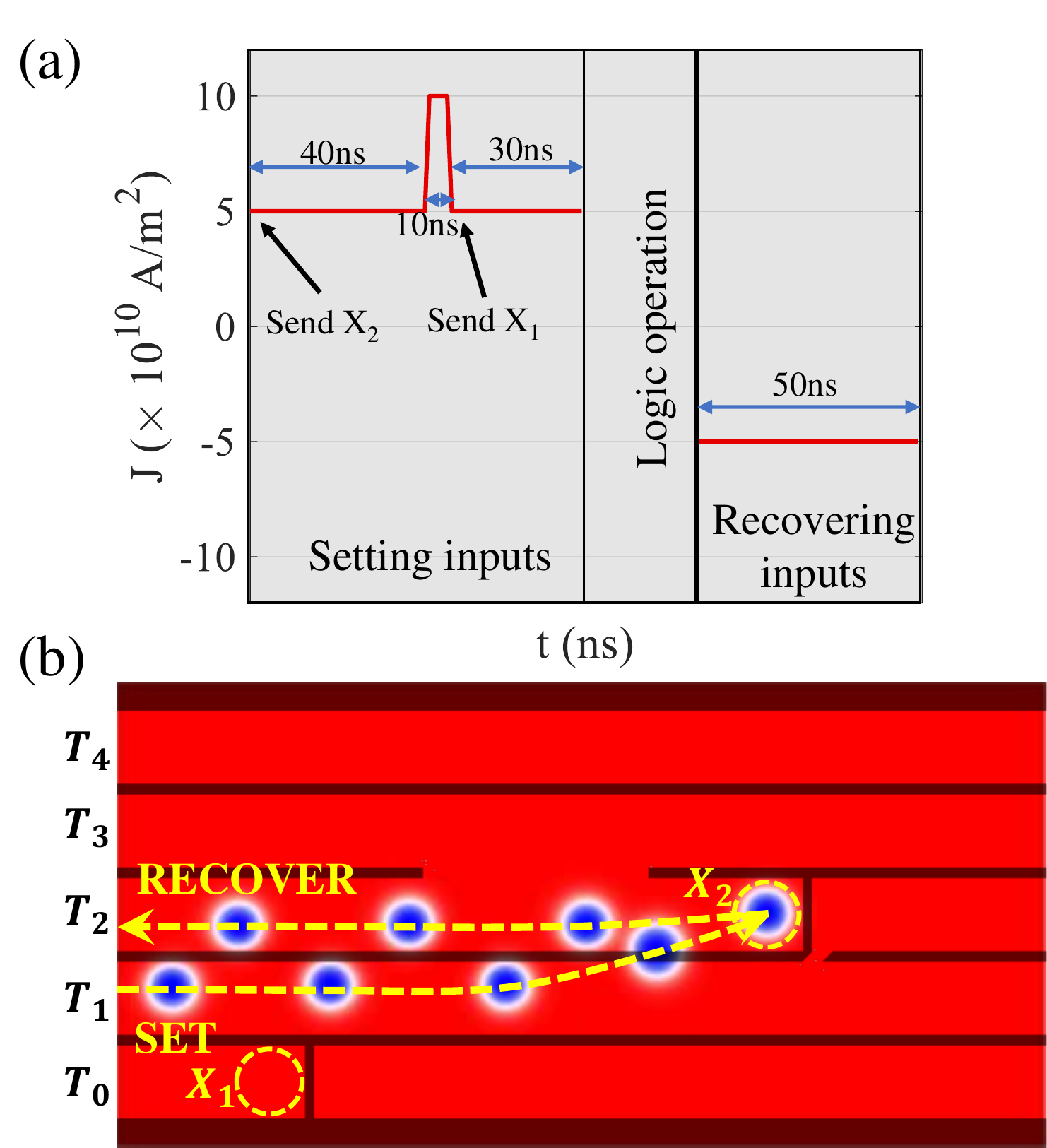}
\caption{\label{fig:setrecover}(a) Set of current pulses used to set the inputs $X_1$ and $X_2$ at their corresponding positions (b) motion of the skyrmion input $X_2$ during a full SET and RECOVER cycle.}
\end{figure}

In order to perform the operations shown in Fig.~\ref{fig:threedmux}, the inputs $X_1$ and $X_2$ have to be set at proper positions. We mention below the protocol followed to achieve this. The input skyrmion $X_2$ is sent through the track $T_1$ at time $t=0$. The injected current is $J=5\times 10^{10}\rm A/m^2$. The skyrmion then moves along the track $T_1$ to the right. At $t=40\rm ns$, we increase the current to $J=10\times 10^{10}\rm A/m^2$ [see Fig.~\ref{fig:setrecover}~(a)]. For this value of current, the skyrmion can pass through the energy barrier. Hence the skyrmion moves from track $T_1$ to the track $T_2$. The current is again reduced back to $J=5\times 10^{10}\rm A/m^2$ at $t=50\rm ns$ once the skyrmion has already entered track $T_2$. At this point, the skyrmion corresponding to input $X_1$ is also injected from the track $T_0$. After another $t=30\rm~ns$ with $J=5\times 10^{10}\rm A/m^2$, both the inputs are placed at their respective positions. We then proceed to perform the logic operations shown in Fig.~\ref{fig:threedmux}. Once the operation is finished, to recover the skyrmion, we simply use a current density with negative polarity $J=-5\times 10^{10}\rm A/m^2$ for $50\rm ns$. The skyrmion input $X_1$ is collected back from the track $T_0$ and the skyrmion input $X_2$ can be collected from the track $T_2$. The motion of the skyrmion input $X_2$ during the entire SET and RECOVER cycle is shown in Fig.~\ref{fig:setrecover}~(b).

\section{\label{sec:cascadefull}Scalability via Cascading}
\begin{figure}
\includegraphics[width=\linewidth]{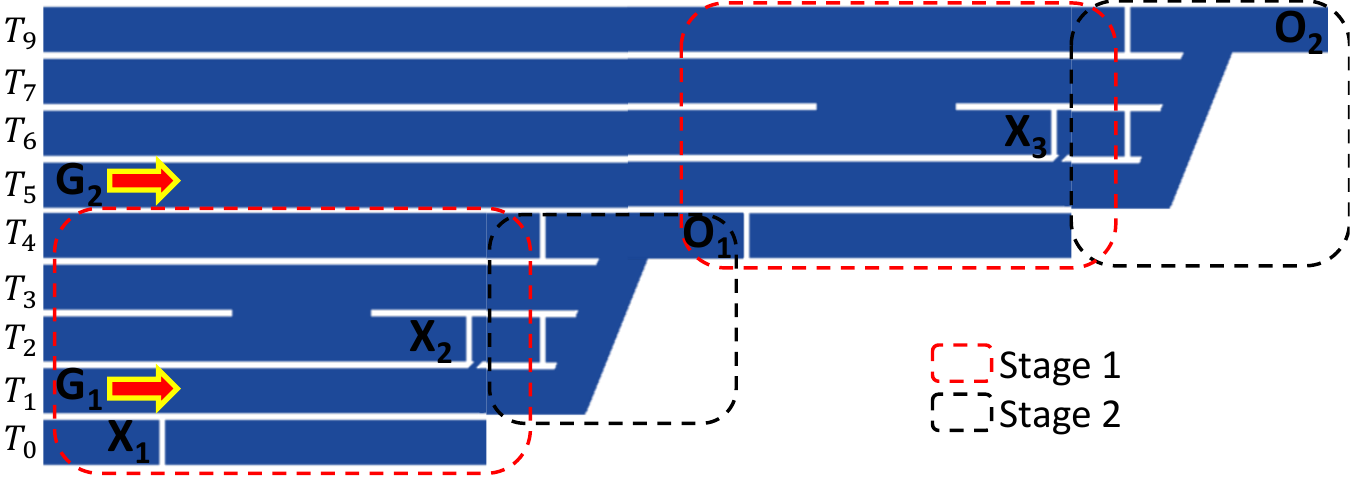}
\caption{\label{fig:scalecascade}Schematic of two cascaded logic gates (each with a Stage 1 and Stage 2 operation) for performing $(X_1 \oplus X_2) \oplus X_3$.}
\end{figure}
The designed logic gates can be combined to make large-scale networks via cascading. In Fig.~\ref{fig:scalecascade}, we show a schematic of two cascaded logic gates each having their own Stage 1 and Stage 2 components. The inputs $X_1$, $X_2$ and $X_3$ will be set by sending them through the tracks $T_0$, $T_1$ and $T_5$, respectively. The fixed input $G_1$ is then sent through the track $T_1$ to perform the first logic operation ($O_1=X_1 \oplus X_2$) followed by sending $G_2$ through the track $T_5$ to perform the second logic operation($O_2=O_1 \oplus X_3$). At the end of the operation, the inputs $X_1$, $X_2$, $X_3$ and outputs $O_1$ (also an input for the second operation), $O_2$ can be retrieved through the tracks $T_0$, $T_2$, $T_6$ and $T_1$, $T_5$, respectively, by sending a negative polarity current.

\section{\label{sec:thermaleffect}Effect of thermal noise}
\begin{figure}[t!]
\includegraphics[width=\linewidth]{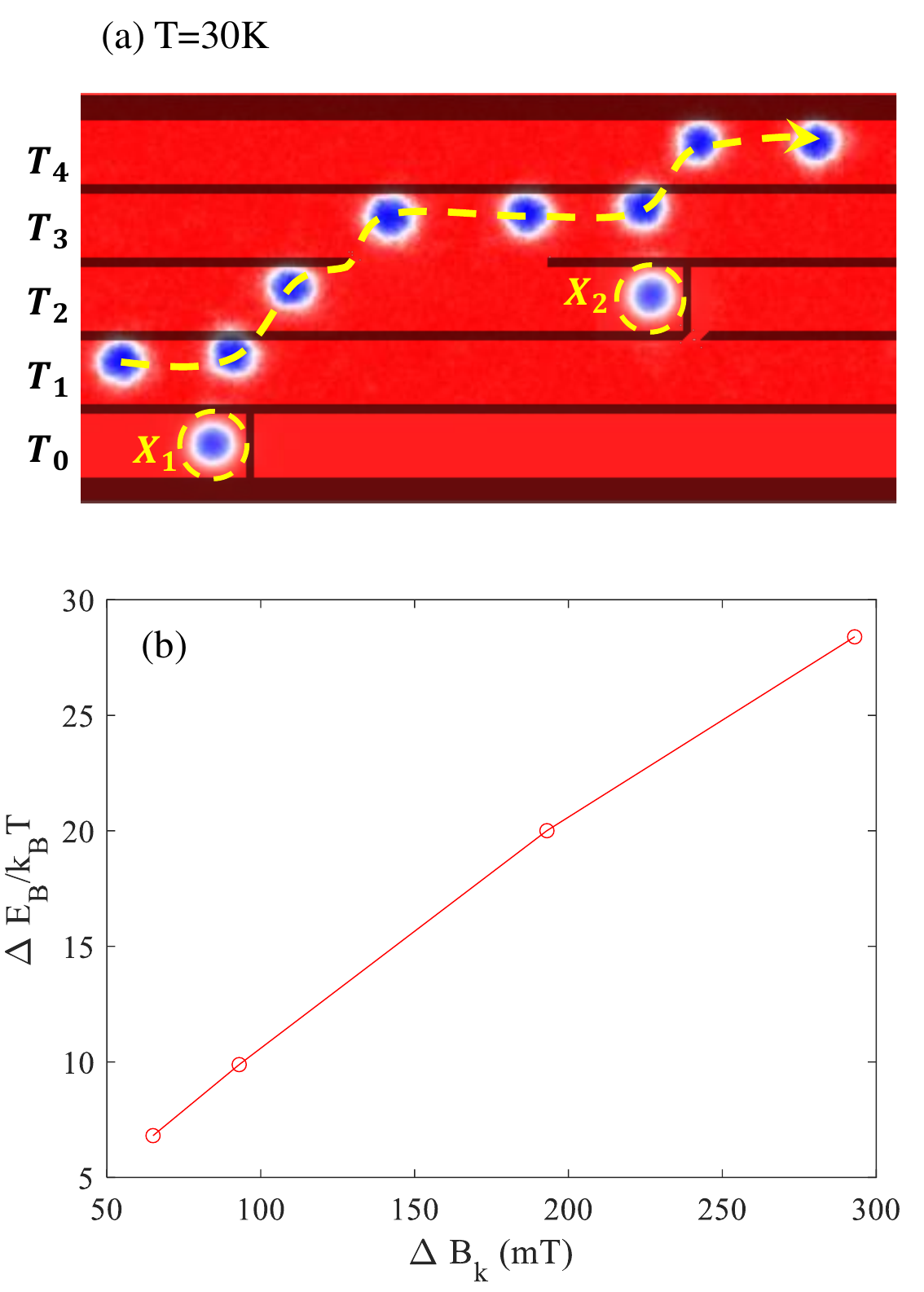}
\caption{\label{fig:temp}(a) Micromagnetic simulation of the  programmable logic \textit{Stage 1} at T=30K showing the trajectory of the skyrmion for input $X_1=1$ and $X_2=1$. (b) Energy barrier as a function of barrier height, $\Delta B_{\rm K}$.}
\end{figure}

In our simulations, we have used the material parameters which were experimentally measured at room temperature. Therefore, we automatically account for the effect of finite temperature on material parameters (reduced $A$, $B_{k}$, $D$ and $M_S$). Nevertheless, in Fig.~\ref{fig:temp}(a), we show the simulation of $Stage 1$ for inputs $X_1=1$ and $X_2=1$ with stochastic noise corresponding to T=30K obtaining a correct logic operation. Note that the quantitative value of temperature used in these simulations may not correspond to real temperatures due to (i) double calculation of thermal effect due to the use of stochastic noise as well as room temperature parameters, (ii) limitations of micromagnetic approach (dependence of thermal field on simulation cell size) and (iii) absence of statistical data (ideally several simulation runs with same parameters but different initial seeds should be performed but this is computationally prohibitive due to long simulation times).

We also calculate the energy of the skyrmion as a function of position from our micromagnetic model. For the parameters used in the main text ($W=12\rm nm$ and $\Delta B_{\rm K}=65\rm mT$), we obtain a barrier energy of $\sim 7\rm k_BT$. Assuming an attempt frequency of 1GHz, the retention time of the skyrmion is $\sim 1\mu s$ which is about 5 times the operation time for the designed logic gate. Moreover, the energy barrier and the retention time can be easily enhanced by increasing $\Delta B_{\rm K}$ as shown in Fig.~\ref{fig:temp}(b).

\section{\label{sec:tolerence}Tolerance w.r.t the variation of $W$ and $\Delta B_{\rm K}$}
\begin{figure}
\includegraphics[width=\linewidth]{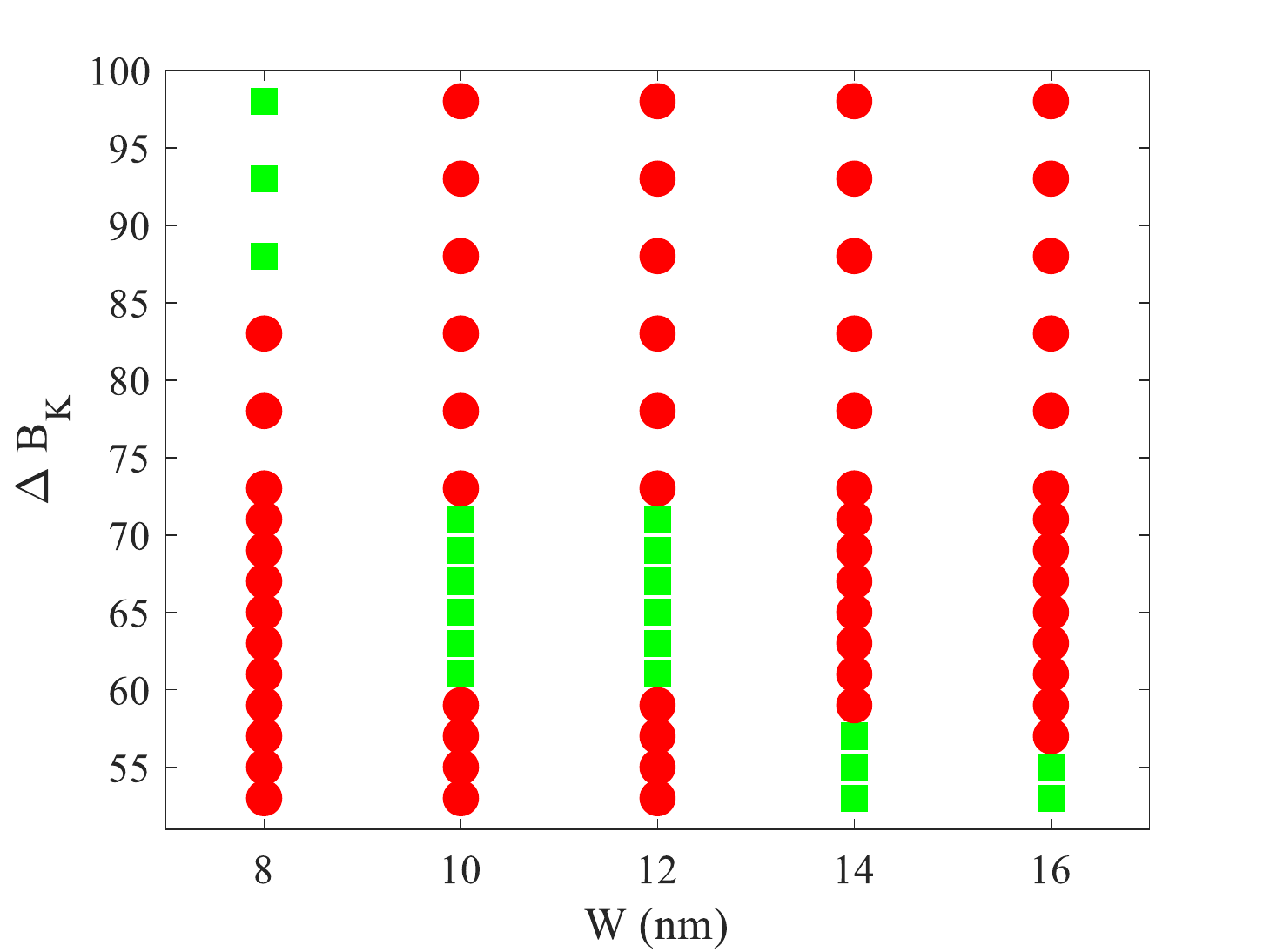}
\caption{\label{fig:tol}Map showing the values of barrier widths $W$ and height $\Delta B_{\rm K}$ for which a correct (green squares) or incorrect (red circles) logic outputs is obtained for DMux Type A.}
\end{figure}

We carried out simulations for DMux Type A logic gate for varying barrier widths $W$ and height $\Delta B_{\rm K}$. The obtained results are shown in Fig.\ref{fig:tol}. The green squares and red circles represent a correct and incorrect logic operation, respectively. In general, the DMux Type A logic gate works for a range of $W$ and $\Delta B_{\rm K}$ values. For fixed $W=12\rm nm$, the logic operation is correctly performed in the range of $\Delta B_{\rm K}=61-71\rm mT$ allowing $\sim15\%$ of variation $\Delta B_{\rm K}$. Higher (lower) barrier width can also be used to design the logic gate, provided that the barrier height is decreased (increased) accordingly

\bibliography{ref}

\begin{thebibliography}{37}%
\makeatletter
\providecommand \@ifxundefined [1]{%
 \@ifx{#1\undefined}
}%
\providecommand \@ifnum [1]{%
 \ifnum #1\expandafter \@firstoftwo
 \else \expandafter \@secondoftwo
 \fi
}%
\providecommand \@ifx [1]{%
 \ifx #1\expandafter \@firstoftwo
 \else \expandafter \@secondoftwo
 \fi
}%
\providecommand \natexlab [1]{#1}%
\providecommand \enquote  [1]{``#1''}%
\providecommand \bibnamefont  [1]{#1}%
\providecommand \bibfnamefont [1]{#1}%
\providecommand \citenamefont [1]{#1}%
\providecommand \href@noop [0]{\@secondoftwo}%
\providecommand \href [0]{\begingroup \@sanitize@url \@href}%
\providecommand \@href[1]{\@@startlink{#1}\@@href}%
\providecommand \@@href[1]{\endgroup#1\@@endlink}%
\providecommand \@sanitize@url [0]{\catcode `\\12\catcode `\$12\catcode
  `\&12\catcode `\#12\catcode `\^12\catcode `\_12\catcode `\%12\relax}%
\providecommand \@@startlink[1]{}%
\providecommand \@@endlink[0]{}%
\providecommand \url  [0]{\begingroup\@sanitize@url \@url }%
\providecommand \@url [1]{\endgroup\@href {#1}{\urlprefix }}%
\providecommand \urlprefix  [0]{URL }%
\providecommand \Eprint [0]{\href }%
\providecommand \doibase [0]{https://doi.org/}%
\providecommand \selectlanguage [0]{\@gobble}%
\providecommand \bibinfo  [0]{\@secondoftwo}%
\providecommand \bibfield  [0]{\@secondoftwo}%
\providecommand \translation [1]{[#1]}%
\providecommand \BibitemOpen [0]{}%
\providecommand \bibitemStop [0]{}%
\providecommand \bibitemNoStop [0]{.\EOS\space}%
\providecommand \EOS [0]{\spacefactor3000\relax}%
\providecommand \BibitemShut  [1]{\csname bibitem#1\endcsname}%
\let\auto@bib@innerbib\@empty
\bibitem [{\citenamefont {Theis}\ and\ \citenamefont
  {Wong}(2017)}]{theis_end_2017}%
  \BibitemOpen
  \bibfield  {author} {\bibinfo {author} {\bibfnamefont {T.~N.}\ \bibnamefont
  {Theis}}\ and\ \bibinfo {author} {\bibfnamefont {H.-S.~P.}\ \bibnamefont
  {Wong}},\ }\bibfield  {title} {\bibinfo {title} {The {End} of {Moore}'s
  {Law}: {A} {New} {Beginning} for {Information} {Technology}},\ }\href
  {https://doi.org/10.1109/MCSE.2017.29} {\bibfield  {journal} {\bibinfo
  {journal} {Computing in Science Engineering}\ }\textbf {\bibinfo {volume}
  {19}},\ \bibinfo {pages} {41} (\bibinfo {year} {2017})}\BibitemShut {NoStop}%
\bibitem [{\citenamefont {Zhirnov}\ \emph {et~al.}(2003)\citenamefont
  {Zhirnov}, \citenamefont {Cavin}, \citenamefont {Hutchby},\ and\
  \citenamefont {Bourianoff}}]{zhirnov_limits_2003}%
  \BibitemOpen
  \bibfield  {author} {\bibinfo {author} {\bibfnamefont {V.}~\bibnamefont
  {Zhirnov}}, \bibinfo {author} {\bibfnamefont {R.}~\bibnamefont {Cavin}},
  \bibinfo {author} {\bibfnamefont {J.}~\bibnamefont {Hutchby}},\ and\ \bibinfo
  {author} {\bibfnamefont {G.}~\bibnamefont {Bourianoff}},\ }\bibfield  {title}
  {\bibinfo {title} {Limits to binary logic switch scaling - a gedanken
  model},\ }\href {https://doi.org/10.1109/JPROC.2003.818324} {\bibfield
  {journal} {\bibinfo  {journal} {Proc. IEEE}\ }\textbf {\bibinfo {volume}
  {91}},\ \bibinfo {pages} {1934} (\bibinfo {year} {2003})}\BibitemShut
  {NoStop}%
\bibitem [{\citenamefont {Nagaosa}\ and\ \citenamefont
  {Tokura}(2013)}]{nagaosa_topological_2013}%
  \BibitemOpen
  \bibfield  {author} {\bibinfo {author} {\bibfnamefont {N.}~\bibnamefont
  {Nagaosa}}\ and\ \bibinfo {author} {\bibfnamefont {Y.}~\bibnamefont
  {Tokura}},\ }\bibfield  {title} {\bibinfo {title} {Topological properties and
  dynamics of magnetic skyrmions},\ }\href
  {https://doi.org/10.1038/nnano.2013.243} {\bibfield  {journal} {\bibinfo
  {journal} {Nat. Nanotech.}\ }\textbf {\bibinfo {volume} {8}},\ \bibinfo
  {pages} {899} (\bibinfo {year} {2013})}\BibitemShut {NoStop}%
\bibitem [{\citenamefont {Fert}\ \emph {et~al.}(2017)\citenamefont {Fert},
  \citenamefont {Reyren},\ and\ \citenamefont {Cros}}]{FertReyrenCros_NRM2017}%
  \BibitemOpen
  \bibfield  {author} {\bibinfo {author} {\bibfnamefont {A.}~\bibnamefont
  {Fert}}, \bibinfo {author} {\bibfnamefont {N.}~\bibnamefont {Reyren}},\ and\
  \bibinfo {author} {\bibfnamefont {V.}~\bibnamefont {Cros}},\ }\bibfield
  {title} {\bibinfo {title} {Magnetic skyrmions: advances in physics and
  potential applications},\ }\href {https://doi.org/10.1038/natrevmats.2017.31}
  {\bibfield  {journal} {\bibinfo  {journal} {Nat.\ Rev.\ Mater.}\ }\textbf
  {\bibinfo {volume} {2}},\ \bibinfo {pages} {17031} (\bibinfo {year}
  {2017})}\BibitemShut {NoStop}%
\bibitem [{\citenamefont {Zhang}\ \emph
  {et~al.}(2015{\natexlab{a}})\citenamefont {Zhang}, \citenamefont {Ezawa},\
  and\ \citenamefont {Zhou}}]{zhang2015magnetic}%
  \BibitemOpen
  \bibfield  {author} {\bibinfo {author} {\bibfnamefont {X.}~\bibnamefont
  {Zhang}}, \bibinfo {author} {\bibfnamefont {M.}~\bibnamefont {Ezawa}},\ and\
  \bibinfo {author} {\bibfnamefont {Y.}~\bibnamefont {Zhou}},\ }\bibfield
  {title} {\bibinfo {title} {Magnetic skyrmion logic gates: conversion,
  duplication and merging of skyrmions},\ }\href@noop {} {\bibfield  {journal}
  {\bibinfo  {journal} {Sci. Rep.}\ }\textbf {\bibinfo {volume} {5}},\ \bibinfo
  {pages} {1} (\bibinfo {year} {2015}{\natexlab{a}})}\BibitemShut {NoStop}%
\bibitem [{\citenamefont {Xing}\ \emph {et~al.}(2016)\citenamefont {Xing},
  \citenamefont {Pong},\ and\ \citenamefont {Zhou}}]{xing2016skyrmion}%
  \BibitemOpen
  \bibfield  {author} {\bibinfo {author} {\bibfnamefont {X.}~\bibnamefont
  {Xing}}, \bibinfo {author} {\bibfnamefont {P.~W.}\ \bibnamefont {Pong}},\
  and\ \bibinfo {author} {\bibfnamefont {Y.}~\bibnamefont {Zhou}},\ }\bibfield
  {title} {\bibinfo {title} {Skyrmion domain wall collision and domain
  wall-gated skyrmion logic},\ }\href@noop {} {\bibfield  {journal} {\bibinfo
  {journal} {Phys.\ Rev.\ B}\ }\textbf {\bibinfo {volume} {94}},\ \bibinfo
  {pages} {054408} (\bibinfo {year} {2016})}\BibitemShut {NoStop}%
\bibitem [{\citenamefont {He}\ \emph {et~al.}(2017)\citenamefont {He},
  \citenamefont {Angizi},\ and\ \citenamefont {Fan}}]{he2017current}%
  \BibitemOpen
  \bibfield  {author} {\bibinfo {author} {\bibfnamefont {Z.}~\bibnamefont
  {He}}, \bibinfo {author} {\bibfnamefont {S.}~\bibnamefont {Angizi}},\ and\
  \bibinfo {author} {\bibfnamefont {D.}~\bibnamefont {Fan}},\ }\bibfield
  {title} {\bibinfo {title} {Current-induced dynamics of multiple skyrmions
  with domain-wall pair and skyrmion-based majority gate design},\ }\href@noop
  {} {\bibfield  {journal} {\bibinfo  {journal} {IEEE Magn. Lett.}\ }\textbf
  {\bibinfo {volume} {8}},\ \bibinfo {pages} {1} (\bibinfo {year}
  {2017})}\BibitemShut {NoStop}%
\bibitem [{\citenamefont {Luo}\ \emph {et~al.}(2018)\citenamefont {Luo},
  \citenamefont {Song}, \citenamefont {Li}, \citenamefont {Zhang},
  \citenamefont {Hong}, \citenamefont {Yang}, \citenamefont {Zou},
  \citenamefont {Xu},\ and\ \citenamefont {You}}]{luo2018reconfigurable}%
  \BibitemOpen
  \bibfield  {author} {\bibinfo {author} {\bibfnamefont {S.}~\bibnamefont
  {Luo}}, \bibinfo {author} {\bibfnamefont {M.}~\bibnamefont {Song}}, \bibinfo
  {author} {\bibfnamefont {X.}~\bibnamefont {Li}}, \bibinfo {author}
  {\bibfnamefont {Y.}~\bibnamefont {Zhang}}, \bibinfo {author} {\bibfnamefont
  {J.}~\bibnamefont {Hong}}, \bibinfo {author} {\bibfnamefont {X.}~\bibnamefont
  {Yang}}, \bibinfo {author} {\bibfnamefont {X.}~\bibnamefont {Zou}}, \bibinfo
  {author} {\bibfnamefont {N.}~\bibnamefont {Xu}},\ and\ \bibinfo {author}
  {\bibfnamefont {L.}~\bibnamefont {You}},\ }\bibfield  {title} {\bibinfo
  {title} {Reconfigurable skyrmion logic gates},\ }\href@noop {} {\bibfield
  {journal} {\bibinfo  {journal} {Nano Lett.}\ }\textbf {\bibinfo {volume}
  {18}},\ \bibinfo {pages} {1180} (\bibinfo {year} {2018})}\BibitemShut
  {NoStop}%
\bibitem [{\citenamefont {Zhang}\ \emph {et~al.}(2019)\citenamefont {Zhang},
  \citenamefont {Zhu}, \citenamefont {Zhang}, \citenamefont {Zhang},
  \citenamefont {Nan}, \citenamefont {Zheng}, \citenamefont {Zhang},\ and\
  \citenamefont {Zhao}}]{zhang2019skyrmion}%
  \BibitemOpen
  \bibfield  {author} {\bibinfo {author} {\bibfnamefont {Z.}~\bibnamefont
  {Zhang}}, \bibinfo {author} {\bibfnamefont {Y.}~\bibnamefont {Zhu}}, \bibinfo
  {author} {\bibfnamefont {Y.}~\bibnamefont {Zhang}}, \bibinfo {author}
  {\bibfnamefont {K.}~\bibnamefont {Zhang}}, \bibinfo {author} {\bibfnamefont
  {J.}~\bibnamefont {Nan}}, \bibinfo {author} {\bibfnamefont {Z.}~\bibnamefont
  {Zheng}}, \bibinfo {author} {\bibfnamefont {Y.}~\bibnamefont {Zhang}},\ and\
  \bibinfo {author} {\bibfnamefont {W.}~\bibnamefont {Zhao}},\ }\bibfield
  {title} {\bibinfo {title} {Skyrmion-based ultra-low power
  electric-field-controlled reconfigurable (super) logic gate},\ }\href@noop {}
  {\bibfield  {journal} {\bibinfo  {journal} {IEEE Electron Device Lett.}\
  }\textbf {\bibinfo {volume} {40}},\ \bibinfo {pages} {1984} (\bibinfo {year}
  {2019})}\BibitemShut {NoStop}%
\bibitem [{\citenamefont {Chauwin}\ \emph {et~al.}(2019)\citenamefont
  {Chauwin}, \citenamefont {Hu}, \citenamefont {Garcia-Sanchez}, \citenamefont
  {Betrabet}, \citenamefont {Paler}, \citenamefont {Moutafis},\ and\
  \citenamefont {Friedman}}]{chauwin2019skyrmion}%
  \BibitemOpen
  \bibfield  {author} {\bibinfo {author} {\bibfnamefont {M.}~\bibnamefont
  {Chauwin}}, \bibinfo {author} {\bibfnamefont {X.}~\bibnamefont {Hu}},
  \bibinfo {author} {\bibfnamefont {F.}~\bibnamefont {Garcia-Sanchez}},
  \bibinfo {author} {\bibfnamefont {N.}~\bibnamefont {Betrabet}}, \bibinfo
  {author} {\bibfnamefont {A.}~\bibnamefont {Paler}}, \bibinfo {author}
  {\bibfnamefont {C.}~\bibnamefont {Moutafis}},\ and\ \bibinfo {author}
  {\bibfnamefont {J.~S.}\ \bibnamefont {Friedman}},\ }\bibfield  {title}
  {\bibinfo {title} {Skyrmion logic system for large-scale reversible
  computation},\ }\href@noop {} {\bibfield  {journal} {\bibinfo  {journal}
  {Phys.\ Rev.\ Appl.}\ }\textbf {\bibinfo {volume} {12}},\ \bibinfo {pages}
  {064053} (\bibinfo {year} {2019})}\BibitemShut {NoStop}%
\bibitem [{\citenamefont {Song}\ \emph {et~al.}(2021)\citenamefont {Song},
  \citenamefont {Park}, \citenamefont {Ko}, \citenamefont {Jang}, \citenamefont
  {Je},\ and\ \citenamefont {Kim}}]{song2021logic}%
  \BibitemOpen
  \bibfield  {author} {\bibinfo {author} {\bibfnamefont {M.}~\bibnamefont
  {Song}}, \bibinfo {author} {\bibfnamefont {M.~G.}\ \bibnamefont {Park}},
  \bibinfo {author} {\bibfnamefont {S.}~\bibnamefont {Ko}}, \bibinfo {author}
  {\bibfnamefont {S.~K.}\ \bibnamefont {Jang}}, \bibinfo {author}
  {\bibfnamefont {M.}~\bibnamefont {Je}},\ and\ \bibinfo {author}
  {\bibfnamefont {K.-J.}\ \bibnamefont {Kim}},\ }\bibfield  {title} {\bibinfo
  {title} {Logic device based on skyrmion annihilation},\ }\href@noop {}
  {\bibfield  {journal} {\bibinfo  {journal} {IEEE Trans. Electron Devices}\ }
  (\bibinfo {year} {2021})}\BibitemShut {NoStop}%
\bibitem [{\citenamefont {Zhang}\ \emph {et~al.}(2020)\citenamefont {Zhang},
  \citenamefont {Zhu}, \citenamefont {Kang}, \citenamefont {Zhang},\ and\
  \citenamefont {Zhao}}]{zhang2020stochastic}%
  \BibitemOpen
  \bibfield  {author} {\bibinfo {author} {\bibfnamefont {H.}~\bibnamefont
  {Zhang}}, \bibinfo {author} {\bibfnamefont {D.}~\bibnamefont {Zhu}}, \bibinfo
  {author} {\bibfnamefont {W.}~\bibnamefont {Kang}}, \bibinfo {author}
  {\bibfnamefont {Y.}~\bibnamefont {Zhang}},\ and\ \bibinfo {author}
  {\bibfnamefont {W.}~\bibnamefont {Zhao}},\ }\bibfield  {title} {\bibinfo
  {title} {Stochastic computing implemented by skyrmionic logic devices},\
  }\href@noop {} {\bibfield  {journal} {\bibinfo  {journal} {Phys.\ Rev.\
  Appl.}\ }\textbf {\bibinfo {volume} {13}},\ \bibinfo {pages} {054049}
  (\bibinfo {year} {2020})}\BibitemShut {NoStop}%
\bibitem [{\citenamefont {Gnoli}\ \emph {et~al.}(2021)\citenamefont {Gnoli},
  \citenamefont {Riente}, \citenamefont {Vacca}, \citenamefont {Ruo~Roch},\
  and\ \citenamefont {Graziano}}]{gnoli2021skyrmion}%
  \BibitemOpen
  \bibfield  {author} {\bibinfo {author} {\bibfnamefont {L.}~\bibnamefont
  {Gnoli}}, \bibinfo {author} {\bibfnamefont {F.}~\bibnamefont {Riente}},
  \bibinfo {author} {\bibfnamefont {M.}~\bibnamefont {Vacca}}, \bibinfo
  {author} {\bibfnamefont {M.}~\bibnamefont {Ruo~Roch}},\ and\ \bibinfo
  {author} {\bibfnamefont {M.}~\bibnamefont {Graziano}},\ }\bibfield  {title}
  {\bibinfo {title} {Skyrmion logic-in-memory architecture for maximum/minimum
  search},\ }\href@noop {} {\bibfield  {journal} {\bibinfo  {journal}
  {Electronics}\ }\textbf {\bibinfo {volume} {10}},\ \bibinfo {pages} {155}
  (\bibinfo {year} {2021})}\BibitemShut {NoStop}%
\bibitem [{\citenamefont {Juge}\ \emph {et~al.}(2021)\citenamefont {Juge},
  \citenamefont {Bairagi}, \citenamefont {Rana}, \citenamefont {Vogel},
  \citenamefont {Sall}, \citenamefont {Mailly}, \citenamefont {Pham},
  \citenamefont {Zhang}, \citenamefont {Sisodia}, \citenamefont {Foerster},
  \citenamefont {Aballe}, \citenamefont {Belmeguenai}, \citenamefont
  {Roussigné}, \citenamefont {Auffret}, \citenamefont {Buda-Prejbeanu},
  \citenamefont {Gaudin}, \citenamefont {Ravelosona},\ and\ \citenamefont
  {Boulle}}]{juge2021helium}%
  \BibitemOpen
  \bibfield  {author} {\bibinfo {author} {\bibfnamefont {R.}~\bibnamefont
  {Juge}}, \bibinfo {author} {\bibfnamefont {K.}~\bibnamefont {Bairagi}},
  \bibinfo {author} {\bibfnamefont {K.~G.}\ \bibnamefont {Rana}}, \bibinfo
  {author} {\bibfnamefont {J.}~\bibnamefont {Vogel}}, \bibinfo {author}
  {\bibfnamefont {M.}~\bibnamefont {Sall}}, \bibinfo {author} {\bibfnamefont
  {D.}~\bibnamefont {Mailly}}, \bibinfo {author} {\bibfnamefont {V.~T.}\
  \bibnamefont {Pham}}, \bibinfo {author} {\bibfnamefont {Q.}~\bibnamefont
  {Zhang}}, \bibinfo {author} {\bibfnamefont {N.}~\bibnamefont {Sisodia}},
  \bibinfo {author} {\bibfnamefont {M.}~\bibnamefont {Foerster}}, \bibinfo
  {author} {\bibfnamefont {L.}~\bibnamefont {Aballe}}, \bibinfo {author}
  {\bibfnamefont {M.}~\bibnamefont {Belmeguenai}}, \bibinfo {author}
  {\bibfnamefont {Y.}~\bibnamefont {Roussigné}}, \bibinfo {author}
  {\bibfnamefont {S.}~\bibnamefont {Auffret}}, \bibinfo {author} {\bibfnamefont
  {L.~D.}\ \bibnamefont {Buda-Prejbeanu}}, \bibinfo {author} {\bibfnamefont
  {G.}~\bibnamefont {Gaudin}}, \bibinfo {author} {\bibfnamefont
  {D.}~\bibnamefont {Ravelosona}},\ and\ \bibinfo {author} {\bibfnamefont
  {O.}~\bibnamefont {Boulle}},\ }\bibfield  {title} {\bibinfo {title} {Helium
  ions put magnetic skyrmions on the track},\ }\href@noop {} {\bibfield
  {journal} {\bibinfo  {journal} {Nano Lett.}\ }\textbf {\bibinfo {volume}
  {21}},\ \bibinfo {pages} {2989} (\bibinfo {year} {2021})}\BibitemShut
  {NoStop}%
\bibitem [{\citenamefont {Song}\ \emph {et~al.}(2020)\citenamefont {Song},
  \citenamefont {Moon}, \citenamefont {Yang}, \citenamefont {Hwang},\ and\
  \citenamefont {Kim}}]{songGuidingDynamicSkyrmions2020}%
  \BibitemOpen
  \bibfield  {author} {\bibinfo {author} {\bibfnamefont {M.}~\bibnamefont
  {Song}}, \bibinfo {author} {\bibfnamefont {K.-W.}\ \bibnamefont {Moon}},
  \bibinfo {author} {\bibfnamefont {S.}~\bibnamefont {Yang}}, \bibinfo {author}
  {\bibfnamefont {C.}~\bibnamefont {Hwang}},\ and\ \bibinfo {author}
  {\bibfnamefont {K.-J.}\ \bibnamefont {Kim}},\ }\bibfield  {title} {\bibinfo
  {title} {Guiding of dynamic skyrmions using chiral magnetic domain wall},\
  }\href {https://doi.org/10.35848/1882-0786/ab8d0b} {\bibfield  {journal}
  {\bibinfo  {journal} {Appl. Phys. Express}\ }\textbf {\bibinfo {volume}
  {13}},\ \bibinfo {pages} {063002} (\bibinfo {year} {2020})}\BibitemShut
  {NoStop}%
\bibitem [{\citenamefont {Purnama}\ \emph {et~al.}(2015)\citenamefont
  {Purnama}, \citenamefont {Gan}, \citenamefont {Wong},\ and\ \citenamefont
  {Lew}}]{purnamaGuidedCurrentinducedSkyrmion2015}%
  \BibitemOpen
  \bibfield  {author} {\bibinfo {author} {\bibfnamefont {I.}~\bibnamefont
  {Purnama}}, \bibinfo {author} {\bibfnamefont {W.~L.}\ \bibnamefont {Gan}},
  \bibinfo {author} {\bibfnamefont {D.~W.}\ \bibnamefont {Wong}},\ and\
  \bibinfo {author} {\bibfnamefont {W.~S.}\ \bibnamefont {Lew}},\ }\bibfield
  {title} {\bibinfo {title} {Guided current-induced skyrmion motion in {{1D}}
  potential well},\ }\href {https://doi.org/10.1038/srep10620} {\bibfield
  {journal} {\bibinfo  {journal} {Sci. Rep.}\ }\textbf {\bibinfo {volume}
  {5}},\ \bibinfo {pages} {10620} (\bibinfo {year} {2015})}\BibitemShut
  {NoStop}%
\bibitem [{\citenamefont {Lai}\ \emph {et~al.}(2017{\natexlab{a}})\citenamefont
  {Lai}, \citenamefont {Zhao}, \citenamefont {Tang}, \citenamefont {Ran},
  \citenamefont {Wu}, \citenamefont {Xia}, \citenamefont {Zhang},\ and\
  \citenamefont {Zhou}}]{laiImprovedRacetrackStructure2017}%
  \BibitemOpen
  \bibfield  {author} {\bibinfo {author} {\bibfnamefont {P.}~\bibnamefont
  {Lai}}, \bibinfo {author} {\bibfnamefont {G.~P.}\ \bibnamefont {Zhao}},
  \bibinfo {author} {\bibfnamefont {H.}~\bibnamefont {Tang}}, \bibinfo {author}
  {\bibfnamefont {N.}~\bibnamefont {Ran}}, \bibinfo {author} {\bibfnamefont
  {S.~Q.}\ \bibnamefont {Wu}}, \bibinfo {author} {\bibfnamefont
  {J.}~\bibnamefont {Xia}}, \bibinfo {author} {\bibfnamefont {X.}~\bibnamefont
  {Zhang}},\ and\ \bibinfo {author} {\bibfnamefont {Y.}~\bibnamefont {Zhou}},\
  }\bibfield  {title} {\bibinfo {title} {An {{Improved Racetrack Structure}}
  for {{Transporting}} a {{Skyrmion}}},\ }\href
  {https://doi.org/10.1038/srep45330} {\bibfield  {journal} {\bibinfo
  {journal} {Sci. Rep.}\ }\textbf {\bibinfo {volume} {7}},\ \bibinfo {pages}
  {45330} (\bibinfo {year} {2017}{\natexlab{a}})}\BibitemShut {NoStop}%
\bibitem [{\citenamefont {Fook}\ \emph {et~al.}(2015)\citenamefont {Fook},
  \citenamefont {Gan}, \citenamefont {Purnama},\ and\ \citenamefont
  {Lew}}]{fookMitigationMagnusForce2015}%
  \BibitemOpen
  \bibfield  {author} {\bibinfo {author} {\bibfnamefont {H.~T.}\ \bibnamefont
  {Fook}}, \bibinfo {author} {\bibfnamefont {W.~L.}\ \bibnamefont {Gan}},
  \bibinfo {author} {\bibfnamefont {I.}~\bibnamefont {Purnama}},\ and\ \bibinfo
  {author} {\bibfnamefont {W.~S.}\ \bibnamefont {Lew}},\ }\bibfield  {title}
  {\bibinfo {title} {Mitigation of {{Magnus Force}} in {{Current}}-{{Induced
  Skyrmion Dynamics}}},\ }\href {https://doi.org/10.1109/TMAG.2015.2433677}
  {\bibfield  {journal} {\bibinfo  {journal} {IEEE Trans. Magn.}\ }\textbf
  {\bibinfo {volume} {51}},\ \bibinfo {pages} {1} (\bibinfo {year}
  {2015})}\BibitemShut {NoStop}%
\bibitem [{\citenamefont {Toscano}\ \emph {et~al.}(2020)\citenamefont
  {Toscano}, \citenamefont {Mendon{\c c}a}, \citenamefont {Miranda},
  \citenamefont {{de Araujo}}, \citenamefont {Sato}, \citenamefont {Coura},\
  and\ \citenamefont {Leonel}}]{toscanoSuppressionSkyrmionHall2020}%
  \BibitemOpen
  \bibfield  {author} {\bibinfo {author} {\bibfnamefont {D.}~\bibnamefont
  {Toscano}}, \bibinfo {author} {\bibfnamefont {J.~P.~A.}\ \bibnamefont
  {Mendon{\c c}a}}, \bibinfo {author} {\bibfnamefont {A.~L.~S.}\ \bibnamefont
  {Miranda}}, \bibinfo {author} {\bibfnamefont {C.~I.~L.}\ \bibnamefont {{de
  Araujo}}}, \bibinfo {author} {\bibfnamefont {F.}~\bibnamefont {Sato}},
  \bibinfo {author} {\bibfnamefont {P.~Z.}\ \bibnamefont {Coura}},\ and\
  \bibinfo {author} {\bibfnamefont {S.~A.}\ \bibnamefont {Leonel}},\ }\bibfield
   {title} {\bibinfo {title} {Suppression of the skyrmion {{Hall}} effect in
  planar nanomagnets by the magnetic properties engineering: {{Skyrmion}}
  transport on nanotracks with magnetic strips},\ }\href
  {https://doi.org/10.1016/j.jmmm.2020.166655} {\bibfield  {journal} {\bibinfo
  {journal} {J.\ Magn.\ Magn.\ Mater.}\ }\textbf {\bibinfo {volume} {504}},\
  \bibinfo {pages} {166655} (\bibinfo {year} {2020})}\BibitemShut {NoStop}%
\bibitem [{\citenamefont {Bhatti}\ and\ \citenamefont
  {Piramanayagam}(2019)}]{bhattiEffectDzyaloshinskiiMoriya2019}%
  \BibitemOpen
  \bibfield  {author} {\bibinfo {author} {\bibfnamefont {S.}~\bibnamefont
  {Bhatti}}\ and\ \bibinfo {author} {\bibfnamefont {S.~N.}\ \bibnamefont
  {Piramanayagam}},\ }\bibfield  {title} {\bibinfo {title} {Effect of
  {{Dzyaloshinskii}}\textendash{{Moriya Interaction Energy Confinement}} on
  {{Current}}-{{Driven Dynamics}} of {{Skyrmions}}},\ }\href
  {https://doi.org/10.1002/pssr.201900090} {\bibfield  {journal} {\bibinfo
  {journal} {Phys.\ Status\ Solidi\ RRL}\ }\textbf {\bibinfo {volume} {13}},\
  \bibinfo {pages} {1900090} (\bibinfo {year} {2019})}\BibitemShut {NoStop}%
\bibitem [{\citenamefont {Loreto}\ \emph {et~al.}(2019)\citenamefont {Loreto},
  \citenamefont {Zhang}, \citenamefont {Zhou}, \citenamefont {Ezawa},
  \citenamefont {Liu},\ and\ \citenamefont {{de
  Araujo}}}]{loretoManipulationMagneticSkyrmions2019}%
  \BibitemOpen
  \bibfield  {author} {\bibinfo {author} {\bibfnamefont {R.~P.}\ \bibnamefont
  {Loreto}}, \bibinfo {author} {\bibfnamefont {X.}~\bibnamefont {Zhang}},
  \bibinfo {author} {\bibfnamefont {Y.}~\bibnamefont {Zhou}}, \bibinfo {author}
  {\bibfnamefont {M.}~\bibnamefont {Ezawa}}, \bibinfo {author} {\bibfnamefont
  {X.}~\bibnamefont {Liu}},\ and\ \bibinfo {author} {\bibfnamefont {C.~I.~L.}\
  \bibnamefont {{de Araujo}}},\ }\bibfield  {title} {\bibinfo {title}
  {Manipulation of magnetic skyrmions in a locally modified synthetic
  antiferromagnetic racetrack},\ }\href
  {https://doi.org/10.1016/j.jmmm.2019.03.030} {\bibfield  {journal} {\bibinfo
  {journal} {J.\ Magn.\ Magn.\ Mater.}\ }\textbf {\bibinfo {volume} {482}},\
  \bibinfo {pages} {155} (\bibinfo {year} {2019})}\BibitemShut {NoStop}%
\bibitem [{\citenamefont {Ang}\ \emph {et~al.}(2019)\citenamefont {Ang},
  \citenamefont {Gan},\ and\ \citenamefont
  {Lew}}]{angBilayerSkyrmionDynamics2019}%
  \BibitemOpen
  \bibfield  {author} {\bibinfo {author} {\bibfnamefont {C.~C.~I.}\
  \bibnamefont {Ang}}, \bibinfo {author} {\bibfnamefont {W.}~\bibnamefont
  {Gan}},\ and\ \bibinfo {author} {\bibfnamefont {W.~S.}\ \bibnamefont {Lew}},\
  }\bibfield  {title} {\bibinfo {title} {Bilayer skyrmion dynamics on a
  magnetic anisotropy gradient},\ }\href
  {https://doi.org/10.1088/1367-2630/ab1171} {\bibfield  {journal} {\bibinfo
  {journal} {New J. Phys.}\ }\textbf {\bibinfo {volume} {21}},\ \bibinfo
  {pages} {043006} (\bibinfo {year} {2019})}\BibitemShut {NoStop}%
\bibitem [{\citenamefont {Iwasaki}\ \emph {et~al.}(2014)\citenamefont
  {Iwasaki}, \citenamefont {Koshibae},\ and\ \citenamefont
  {Nagaosa}}]{iwasakiColossalSpinTransfer2014}%
  \BibitemOpen
  \bibfield  {author} {\bibinfo {author} {\bibfnamefont {J.}~\bibnamefont
  {Iwasaki}}, \bibinfo {author} {\bibfnamefont {W.}~\bibnamefont {Koshibae}},\
  and\ \bibinfo {author} {\bibfnamefont {N.}~\bibnamefont {Nagaosa}},\
  }\bibfield  {title} {\bibinfo {title} {Colossal {{Spin Transfer Torque
  Effect}} on {{Skyrmion}} along the {{Edge}}},\ }\href
  {https://doi.org/10.1021/nl501379k} {\bibfield  {journal} {\bibinfo
  {journal} {Nano Lett.}\ }\textbf {\bibinfo {volume} {14}},\ \bibinfo {pages}
  {4432} (\bibinfo {year} {2014})}\BibitemShut {NoStop}%
\bibitem [{\citenamefont {Lai}\ \emph {et~al.}(2017{\natexlab{b}})\citenamefont
  {Lai}, \citenamefont {Zhao}, \citenamefont {Morvan}, \citenamefont {Wu},\
  and\ \citenamefont {Ran}}]{laiMotionSkyrmionsWellSeparated2017}%
  \BibitemOpen
  \bibfield  {author} {\bibinfo {author} {\bibfnamefont {P.}~\bibnamefont
  {Lai}}, \bibinfo {author} {\bibfnamefont {G.~P.}\ \bibnamefont {Zhao}},
  \bibinfo {author} {\bibfnamefont {F.~J.}\ \bibnamefont {Morvan}}, \bibinfo
  {author} {\bibfnamefont {S.~Q.}\ \bibnamefont {Wu}},\ and\ \bibinfo {author}
  {\bibfnamefont {N.}~\bibnamefont {Ran}},\ }\bibfield  {title} {\bibinfo
  {title} {Motion of {{Skyrmions}} in {{Well}}-{{Separated Two}}-{{Lane
  Racetracks}}},\ }\href {https://doi.org/10.1142/S2010324717400069} {\bibfield
   {journal} {\bibinfo  {journal} {SPIN}\ }\textbf {\bibinfo {volume} {07}},\
  \bibinfo {pages} {1740006} (\bibinfo {year}
  {2017}{\natexlab{b}})}\BibitemShut {NoStop}%
\bibitem [{\citenamefont {Zhang}\ \emph
  {et~al.}(2015{\natexlab{b}})\citenamefont {Zhang}, \citenamefont {Zhao},
  \citenamefont {Fangohr}, \citenamefont {Liu}, \citenamefont {Xia},
  \citenamefont {Xia},\ and\ \citenamefont
  {Morvan}}]{zhangSkyrmionskyrmionSkyrmionedgeRepulsions2015}%
  \BibitemOpen
  \bibfield  {author} {\bibinfo {author} {\bibfnamefont {X.}~\bibnamefont
  {Zhang}}, \bibinfo {author} {\bibfnamefont {G.~P.}\ \bibnamefont {Zhao}},
  \bibinfo {author} {\bibfnamefont {H.}~\bibnamefont {Fangohr}}, \bibinfo
  {author} {\bibfnamefont {J.~P.}\ \bibnamefont {Liu}}, \bibinfo {author}
  {\bibfnamefont {W.~X.}\ \bibnamefont {Xia}}, \bibinfo {author} {\bibfnamefont
  {J.}~\bibnamefont {Xia}},\ and\ \bibinfo {author} {\bibfnamefont {F.~J.}\
  \bibnamefont {Morvan}},\ }\bibfield  {title} {\bibinfo {title}
  {Skyrmion-skyrmion and skyrmion-edge repulsions in skyrmion-based racetrack
  memory},\ }\href {https://doi.org/10.1038/srep07643} {\bibfield  {journal}
  {\bibinfo  {journal} {Sci. Rep.}\ }\textbf {\bibinfo {volume} {5}},\ \bibinfo
  {pages} {7643} (\bibinfo {year} {2015}{\natexlab{b}})}\BibitemShut {NoStop}%
\bibitem [{\citenamefont {Yan}\ \emph {et~al.}(2020)\citenamefont {Yan},
  \citenamefont {Liu}, \citenamefont {Guang}, \citenamefont {Feng},
  \citenamefont {Lake}, \citenamefont {Yu},\ and\ \citenamefont
  {Han}}]{yanRobustSkyrmionShift2020}%
  \BibitemOpen
  \bibfield  {author} {\bibinfo {author} {\bibfnamefont {Z.}~\bibnamefont
  {Yan}}, \bibinfo {author} {\bibfnamefont {Y.}~\bibnamefont {Liu}}, \bibinfo
  {author} {\bibfnamefont {Y.}~\bibnamefont {Guang}}, \bibinfo {author}
  {\bibfnamefont {J.}~\bibnamefont {Feng}}, \bibinfo {author} {\bibfnamefont
  {R.}~\bibnamefont {Lake}}, \bibinfo {author} {\bibfnamefont {G.}~\bibnamefont
  {Yu}},\ and\ \bibinfo {author} {\bibfnamefont {X.}~\bibnamefont {Han}},\
  }\bibfield  {title} {\bibinfo {title} {Robust {{Skyrmion Shift Device Through
  Engineering}} the {{Local Exchange}}-{{Bias Field}}},\ }\href
  {https://doi.org/10.1103/PhysRevApplied.14.044008} {\bibfield  {journal}
  {\bibinfo  {journal} {Phys. Rev. Applied}\ }\textbf {\bibinfo {volume}
  {14}},\ \bibinfo {pages} {044008} (\bibinfo {year} {2020})}\BibitemShut
  {NoStop}%
\bibitem [{\citenamefont {Ang}\ \emph {et~al.}(2020)\citenamefont {Ang},
  \citenamefont {Gan}, \citenamefont {Wong},\ and\ \citenamefont
  {Lew}}]{angElectricalControlSkyrmion2020}%
  \BibitemOpen
  \bibfield  {author} {\bibinfo {author} {\bibfnamefont {C.~C.~I.}\
  \bibnamefont {Ang}}, \bibinfo {author} {\bibfnamefont {W.}~\bibnamefont
  {Gan}}, \bibinfo {author} {\bibfnamefont {G.~D.~H.}\ \bibnamefont {Wong}},\
  and\ \bibinfo {author} {\bibfnamefont {W.~S.}\ \bibnamefont {Lew}},\
  }\bibfield  {title} {\bibinfo {title} {Electrical {{Control}} of {{Skyrmion
  Density}} via {{Skyrmion}}-{{Stripe Transformation}}},\ }\href
  {https://doi.org/10.1103/PhysRevApplied.14.054048} {\bibfield  {journal}
  {\bibinfo  {journal} {Phys. Rev. Applied}\ }\textbf {\bibinfo {volume}
  {14}},\ \bibinfo {pages} {054048} (\bibinfo {year} {2020})}\BibitemShut
  {NoStop}%
\bibitem [{\citenamefont {Sapozhnikov}\ \emph {et~al.}(2016)\citenamefont
  {Sapozhnikov}, \citenamefont {Vdovichev}, \citenamefont {Ermolaeva},
  \citenamefont {Gusev}, \citenamefont {Fraerman}, \citenamefont {Gusev},\ and\
  \citenamefont {Petrov}}]{sapozhnikovArtificialDenseLattice2016}%
  \BibitemOpen
  \bibfield  {author} {\bibinfo {author} {\bibfnamefont {M.~V.}\ \bibnamefont
  {Sapozhnikov}}, \bibinfo {author} {\bibfnamefont {S.~N.}\ \bibnamefont
  {Vdovichev}}, \bibinfo {author} {\bibfnamefont {O.~L.}\ \bibnamefont
  {Ermolaeva}}, \bibinfo {author} {\bibfnamefont {N.~S.}\ \bibnamefont
  {Gusev}}, \bibinfo {author} {\bibfnamefont {A.~A.}\ \bibnamefont {Fraerman}},
  \bibinfo {author} {\bibfnamefont {S.~A.}\ \bibnamefont {Gusev}},\ and\
  \bibinfo {author} {\bibfnamefont {Y.~V.}\ \bibnamefont {Petrov}},\ }\bibfield
   {title} {\bibinfo {title} {Artificial dense lattice of magnetic bubbles},\
  }\href {https://doi.org/10.1063/1.4958300} {\bibfield  {journal} {\bibinfo
  {journal} {Appl. Phys. Lett.}\ }\textbf {\bibinfo {volume} {109}},\ \bibinfo
  {pages} {042406} (\bibinfo {year} {2016})}\BibitemShut {NoStop}%
\bibitem [{\citenamefont {Toscano}\ \emph {et~al.}(2019)\citenamefont
  {Toscano}, \citenamefont {Leonel}, \citenamefont {Coura},\ and\ \citenamefont
  {Sato}}]{toscanoBuildingTrapsSkyrmions2019}%
  \BibitemOpen
  \bibfield  {author} {\bibinfo {author} {\bibfnamefont {D.}~\bibnamefont
  {Toscano}}, \bibinfo {author} {\bibfnamefont {S.~A.}\ \bibnamefont {Leonel}},
  \bibinfo {author} {\bibfnamefont {P.~Z.}\ \bibnamefont {Coura}},\ and\
  \bibinfo {author} {\bibfnamefont {F.}~\bibnamefont {Sato}},\ }\bibfield
  {title} {\bibinfo {title} {Building traps for skyrmions by the incorporation
  of magnetic defects into nanomagnets: {{Pinning}} and scattering traps by
  magnetic properties engineering},\ }\href
  {https://doi.org/10.1016/j.jmmm.2019.02.075} {\bibfield  {journal} {\bibinfo
  {journal} {J.\ Magn.\ Magn.\ Mater.}\ }\textbf {\bibinfo {volume} {480}},\
  \bibinfo {pages} {171} (\bibinfo {year} {2019})}\BibitemShut {NoStop}%
\bibitem [{\citenamefont {Sapozhnikov}\ \emph {et~al.}(2018)\citenamefont
  {Sapozhnikov}, \citenamefont {Ermolaeva}, \citenamefont {Skorokhodov},
  \citenamefont {Gusev},\ and\ \citenamefont
  {Drozdov}}]{sapozhnikovMagneticSkyrmionsThicknessModulated2018}%
  \BibitemOpen
  \bibfield  {author} {\bibinfo {author} {\bibfnamefont {M.~V.}\ \bibnamefont
  {Sapozhnikov}}, \bibinfo {author} {\bibfnamefont {O.~V.}\ \bibnamefont
  {Ermolaeva}}, \bibinfo {author} {\bibfnamefont {E.~V.}\ \bibnamefont
  {Skorokhodov}}, \bibinfo {author} {\bibfnamefont {N.~S.}\ \bibnamefont
  {Gusev}},\ and\ \bibinfo {author} {\bibfnamefont {M.~N.}\ \bibnamefont
  {Drozdov}},\ }\bibfield  {title} {\bibinfo {title} {Magnetic {{Skyrmions}} in
  {{Thickness}}-{{Modulated Films}}},\ }\href
  {https://doi.org/10.1134/S0021364018060115} {\bibfield  {journal} {\bibinfo
  {journal} {Jetp Lett.}\ }\textbf {\bibinfo {volume} {107}},\ \bibinfo {pages}
  {364} (\bibinfo {year} {2018})}\BibitemShut {NoStop}%
\bibitem [{\citenamefont {Pinna}\ \emph {et~al.}(2018)\citenamefont {Pinna},
  \citenamefont {Abreu~Araujo}, \citenamefont {Kim}, \citenamefont {Cros},
  \citenamefont {Querlioz}, \citenamefont {Bessiere}, \citenamefont {Droulez},\
  and\ \citenamefont {Grollier}}]{pinna_skyrmion_2017}%
  \BibitemOpen
  \bibfield  {author} {\bibinfo {author} {\bibfnamefont {D.}~\bibnamefont
  {Pinna}}, \bibinfo {author} {\bibfnamefont {F.}~\bibnamefont {Abreu~Araujo}},
  \bibinfo {author} {\bibfnamefont {J.-V.}\ \bibnamefont {Kim}}, \bibinfo
  {author} {\bibfnamefont {V.}~\bibnamefont {Cros}}, \bibinfo {author}
  {\bibfnamefont {D.}~\bibnamefont {Querlioz}}, \bibinfo {author}
  {\bibfnamefont {P.}~\bibnamefont {Bessiere}}, \bibinfo {author}
  {\bibfnamefont {J.}~\bibnamefont {Droulez}},\ and\ \bibinfo {author}
  {\bibfnamefont {J.}~\bibnamefont {Grollier}},\ }\bibfield  {title} {\bibinfo
  {title} {Skyrmion gas manipulation for probabilistic computing},\ }\href
  {https://doi.org/10.1103/PhysRevApplied.9.064018} {\bibfield  {journal}
  {\bibinfo  {journal} {Phys. Rev. Applied}\ }\textbf {\bibinfo {volume} {9}},\
  \bibinfo {pages} {064018} (\bibinfo {year} {2018})}\BibitemShut {NoStop}%
\bibitem [{\citenamefont {Chappert}\ \emph {et~al.}(1998)\citenamefont
  {Chappert}, \citenamefont {Bernas}, \citenamefont {Ferré}, \citenamefont
  {Kottler}, \citenamefont {Jamet}, \citenamefont {Chen}, \citenamefont
  {Cambril}, \citenamefont {Devolder}, \citenamefont {Rousseaux}, \citenamefont
  {Mathet},\ and\ \citenamefont {Launois}}]{chappert_planar_1998}%
  \BibitemOpen
  \bibfield  {author} {\bibinfo {author} {\bibfnamefont {C.}~\bibnamefont
  {Chappert}}, \bibinfo {author} {\bibfnamefont {H.}~\bibnamefont {Bernas}},
  \bibinfo {author} {\bibfnamefont {J.}~\bibnamefont {Ferré}}, \bibinfo
  {author} {\bibfnamefont {V.}~\bibnamefont {Kottler}}, \bibinfo {author}
  {\bibfnamefont {J.-P.}\ \bibnamefont {Jamet}}, \bibinfo {author}
  {\bibfnamefont {Y.}~\bibnamefont {Chen}}, \bibinfo {author} {\bibfnamefont
  {E.}~\bibnamefont {Cambril}}, \bibinfo {author} {\bibfnamefont
  {T.}~\bibnamefont {Devolder}}, \bibinfo {author} {\bibfnamefont
  {F.}~\bibnamefont {Rousseaux}}, \bibinfo {author} {\bibfnamefont
  {V.}~\bibnamefont {Mathet}},\ and\ \bibinfo {author} {\bibfnamefont
  {H.}~\bibnamefont {Launois}},\ }\bibfield  {title} {\bibinfo {title} {Planar
  {Patterned} {Magnetic} {Media} {Obtained} by {Ion} {Irradiation}},\ }\href
  {https://doi.org/10.1126/science.280.5371.1919} {\bibfield  {journal}
  {\bibinfo  {journal} {Science}\ }\textbf {\bibinfo {volume} {280}},\ \bibinfo
  {pages} {1919} (\bibinfo {year} {1998})}\BibitemShut {NoStop}%
\bibitem [{\citenamefont {Menezes}\ \emph {et~al.}(2019)\citenamefont
  {Menezes}, \citenamefont {Mulkers}, \citenamefont {Silva},\ and\
  \citenamefont {Milošević}}]{menezes_deflection_2019}%
  \BibitemOpen
  \bibfield  {author} {\bibinfo {author} {\bibfnamefont {R.~M.}\ \bibnamefont
  {Menezes}}, \bibinfo {author} {\bibfnamefont {J.}~\bibnamefont {Mulkers}},
  \bibinfo {author} {\bibfnamefont {C.~C. d.~S.}\ \bibnamefont {Silva}},\ and\
  \bibinfo {author} {\bibfnamefont {M.~V.}\ \bibnamefont {Milošević}},\
  }\bibfield  {title} {\bibinfo {title} {Deflection of ferromagnetic and
  antiferromagnetic skyrmions at heterochiral interfaces},\ }\href
  {https://doi.org/10.1103/PhysRevB.99.104409} {\bibfield  {journal} {\bibinfo
  {journal} {Phys.\ Rev.\ B}\ }\textbf {\bibinfo {volume} {99}},\ \bibinfo
  {pages} {104409} (\bibinfo {year} {2019})}\BibitemShut {NoStop}%
\bibitem [{\citenamefont {Castell-Queralt}\ \emph {et~al.}(2019)\citenamefont
  {Castell-Queralt}, \citenamefont {Gonz{\'a}lez-G{\'o}mez}, \citenamefont
  {Del-Valle}, \citenamefont {Sanchez},\ and\ \citenamefont
  {Navau}}]{castell2019accelerating}%
  \BibitemOpen
  \bibfield  {author} {\bibinfo {author} {\bibfnamefont {J.}~\bibnamefont
  {Castell-Queralt}}, \bibinfo {author} {\bibfnamefont {L.}~\bibnamefont
  {Gonz{\'a}lez-G{\'o}mez}}, \bibinfo {author} {\bibfnamefont {N.}~\bibnamefont
  {Del-Valle}}, \bibinfo {author} {\bibfnamefont {A.}~\bibnamefont {Sanchez}},\
  and\ \bibinfo {author} {\bibfnamefont {C.}~\bibnamefont {Navau}},\ }\bibfield
   {title} {\bibinfo {title} {Accelerating, guiding, and compressing skyrmions
  by defect rails},\ }\href@noop {} {\bibfield  {journal} {\bibinfo  {journal}
  {Nanoscale}\ }\textbf {\bibinfo {volume} {11}},\ \bibinfo {pages} {12589}
  (\bibinfo {year} {2019})}\BibitemShut {NoStop}%
\bibitem [{\citenamefont {Lewis}\ \emph {et~al.}(2019)\citenamefont {Lewis},
  \citenamefont {Hunt}, \citenamefont {DeRose}, \citenamefont {Alty},
  \citenamefont {Li}, \citenamefont {Wertheim}, \citenamefont {De~Rose},
  \citenamefont {Timco}, \citenamefont {Scherer}, \citenamefont {Yeates} \emph
  {et~al.}}]{lewis2019plasma}%
  \BibitemOpen
  \bibfield  {author} {\bibinfo {author} {\bibfnamefont {S.~M.}\ \bibnamefont
  {Lewis}}, \bibinfo {author} {\bibfnamefont {M.~S.}\ \bibnamefont {Hunt}},
  \bibinfo {author} {\bibfnamefont {G.~A.}\ \bibnamefont {DeRose}}, \bibinfo
  {author} {\bibfnamefont {H.~R.}\ \bibnamefont {Alty}}, \bibinfo {author}
  {\bibfnamefont {J.}~\bibnamefont {Li}}, \bibinfo {author} {\bibfnamefont
  {A.}~\bibnamefont {Wertheim}}, \bibinfo {author} {\bibfnamefont
  {L.}~\bibnamefont {De~Rose}}, \bibinfo {author} {\bibfnamefont {G.~A.}\
  \bibnamefont {Timco}}, \bibinfo {author} {\bibfnamefont {A.}~\bibnamefont
  {Scherer}}, \bibinfo {author} {\bibfnamefont {S.~G.}\ \bibnamefont {Yeates}},
  \emph {et~al.},\ }\bibfield  {title} {\bibinfo {title} {Plasma-etched pattern
  transfer of sub-10 nm structures using a metal--organic resist and helium ion
  beam lithography},\ }\href@noop {} {\bibfield  {journal} {\bibinfo  {journal}
  {Nano Lett.}\ }\textbf {\bibinfo {volume} {19}},\ \bibinfo {pages} {6043}
  (\bibinfo {year} {2019})}\BibitemShut {NoStop}%
\bibitem [{\citenamefont {Kluge}(1972)}]{kluge1972computation}%
  \BibitemOpen
  \bibfield  {author} {\bibinfo {author} {\bibfnamefont {W.}~\bibnamefont
  {Kluge}},\ }\bibfield  {title} {\bibinfo {title} {Computation of switching
  functions using input-pattern-conserving magnetic-bubble manipulations},\
  }\href@noop {} {\bibfield  {journal} {\bibinfo  {journal} {Electronics
  Letters}\ }\textbf {\bibinfo {volume} {8}},\ \bibinfo {pages} {313} (\bibinfo
  {year} {1972})}\BibitemShut {NoStop}%
\bibitem [{\citenamefont {{Vansteenkiste}}\ \emph {et~al.}(2014)\citenamefont
  {{Vansteenkiste}}, \citenamefont {{Leliaert}}, \citenamefont {{Dvornik}},
  \citenamefont {{Helsen}}, \citenamefont {{Garcia-Sanchez}},\ and\
  \citenamefont {{Van Waeyenberge}}}]{VansteenkisteWaeyenberge_AIPadv2014}%
  \BibitemOpen
  \bibfield  {author} {\bibinfo {author} {\bibfnamefont {A.}~\bibnamefont
  {{Vansteenkiste}}}, \bibinfo {author} {\bibfnamefont {J.}~\bibnamefont
  {{Leliaert}}}, \bibinfo {author} {\bibfnamefont {M.}~\bibnamefont
  {{Dvornik}}}, \bibinfo {author} {\bibfnamefont {M.}~\bibnamefont {{Helsen}}},
  \bibinfo {author} {\bibfnamefont {F.}~\bibnamefont {{Garcia-Sanchez}}},\ and\
  \bibinfo {author} {\bibfnamefont {B.}~\bibnamefont {{Van Waeyenberge}}},\
  }\bibfield  {title} {\bibinfo {title} {{The design and verification of
  MuMax3}},\ }\href {https://doi.org/10.1063/1.4899186} {\bibfield  {journal}
  {\bibinfo  {journal} {AIP Adv.}\ }\textbf {\bibinfo {volume} {4}},\ \bibinfo
  {eid} {107133} (\bibinfo {year} {2014})}\BibitemShut {NoStop}%
\end{thebibliography}%

\end{document}